\documentclass[aps,pra,,notitlepage]{revtex4}
\usepackage{latexsym}
\usepackage{amsmath}
\usepackage{amssymb}
\usepackage{bm}
\usepackage{wasysym}
\usepackage[dvips]{color}

\usepackage{graphicx}
\usepackage{graphicx}
\usepackage{subfigure}
\usepackage{epsfig}

\newcommand{\hide}[1]{}

\begin{document}

\title{Open Quantum System Stochastic Dynamics with and without the 
RWA}

\author{Y. B. Band}
\affiliation{Department of Chemistry, Department of Physics,
Department of Electro-Optics, and the Ilse Katz Center for
Nano-Science, Ben-Gurion University, Beer-Sheva 84105, Israel}


\begin{abstract}
We study the dynamics of a two-level quantum system interacting with a
single frequency electromagnetic field and a stochastic magnetic field, 
with and without making the rotating wave approximation (RWA).
The transformation to the rotating frame does not commute with the
stochastic Hamiltonian if the stochastic field has nonvanishing
components in the transverse direction, hence, applying the RWA
requires transformation of the stochastic terms in the Hamiltonian.
For Gaussian white noise, the master equation is derived from the
stochastic Schr\"{o}dinger-Langevin equations, with and without the
RWA. With the RWA, the master equation for the density matrix has 
Lindblad terms with coefficients that are time-dependent (i.e., the
master equation is time-local).
An approximate analytic expression for the density matrix is obtained
with the RWA. For Ornstein--Uhlenbeck noise, as well as other types of
colored noise, in contradistinction to the Gaussian white noise case,
the non-commutation of the RWA transformation and the noise
Hamiltonian can significantly affect the RWA dynamics when $\omega
\tau_{\mathrm{corr}} \gtrapprox 1$, where $\omega$ is the
electromagnetic field frequency and $\tau_{\mathrm{corr}}$ is the
stochastic magnetic field correlation time.
\end{abstract}


\maketitle

\section{Introduction} \label{Sec:Intro}

One of the basic quantum processes studied in physics is the
two-level system driven by an electromagnetic field.  At least six
Nobel prizes were awarded for work on such processes: I. I. Rabi, for
the resonance method applied to molecules and NMR, F. Bloch and E. M.
Purcell for their development of new methods for NMR, C. W. Townes,
N. G. Basov, and A. M. Prokhorov for masers, lasers and quantum
optics, A. Kastler for optical pumping, N. F. Ramsey for the separated
oscillatory fields method and its use in atom clocks, and S. Haroche
and D. J. Wineland for developing methods for observing individual
quantum particles without destroying them.  But quantum systems are
never isolated; they interact with their environment, and this gives
rise to perturbations that can strongly affect their behavior.  Such
interaction affects all the phenomena enumerated above, as well as
other phenomena including dephasing in metals \cite{Aleiner_01},
nuclear-spin-dependent ground-state dephasing in diamond
nitrogen-vacency centers \cite{NV}, broadening and shift of atomic
clock transitions \cite{ChangYeLukin_04}, and decoherence in quantum
information processes \cite{qc}.

The dynamics of a quantum system coupled to an environment (a bath) is
often treated in terms of the reduced density matrix of the system
obtained by tracing out the bath degrees of freedom in the state of
the system plus bath \cite{master_eq}.  Upon assuming that the initial
density matrix is of a product state form, making the Born-Markov
approximation and the rotating wave approximation (RWA) \cite{master_eq}, 
the resulting master equation for the 
reduced density matrix is of Lindblad form \cite{Lindblad_76}.  An 
alternative treatment models the coupling of the quantum system and the 
bath by introducing stochastic fields that act on the system, where the
stochastic fields are generated by a complex environment \cite{vanKampenBook, 
Gardiner, Fleming_12b, Band}.  The statistical properties of
the stochastic fields are determined by the properties of the
environment.  The environment can sometimes be modeled as a ensemble of 
approximately independent fluctuating fields in steady state (e.g.,
in thermal equilibrium).  In this approximation, the resultant stochastic 
field felt by the system is a superposition of a large number of 
components.  Due to the central limit theorem \cite{clt}, the stochastic
fields can be represented by Gaussian, stationary stochastic processes
which are completely specified by their first two moments.  Moreover,
if the timescales of the bath are small compared to those of the
system, the stochastic processes can be approximated to be Gaussian
white noise.  The averaged (over stochastic realizations) quantities
obtained for Gaussian white noise are equivalent to the averages obtained
using a Lindblad master equation approach \cite{vanKampenBook} (see
Sec.~\ref{Sec:Master_Eqs}).  The stochastic process method used here
is called the {\em Schr\"{o}dinger-Langevin stochastic differential
equation formalism} (SLSDE) \cite{vanKampenBook}.  In principle, the bath
could be affected by the system (back-action).  This back-action would
modify the properties of the noise felt by the system and effectively
appear as a self-interaction mediated by the environment.  However, if
the perturbation caused to the environment by the quantum system is
weak, back-action can be neglected \cite{vanKampenBook, Gardiner}.  The 
neglect of back-action is similar to one of the approximations (the Born
approximation) made in the Born-Markov approximation of
the master equation approach.  Neglect of back-action is called, in
the context of the SLSDE formalism, the {\it external noise approximation} 
\cite{vanKampenBook}.

Let us explicitly consider a two-level system, e.g., a spin 1/2
particle.  The system interacts with a constant magnetic field, whose
direction can be taken, without loss of generality, to be along the
$z$ axis, an electromagnetic field with frequency $\omega$, and a
stochastic magnetic field, which can be viewed as being due to
interaction with a bath of other particles having magnetic dipole
moments.  The deterministic part of the Hamiltonian for the system
can be written as \cite{semiclass}
\begin{equation} \label{Eq:H}
    H(t) = \hbar \begin{pmatrix} \frac{\delta}{2} & \Omega \sin
    \omega t \\
    \Omega \sin \omega t & - \frac{\delta}{2} \end{pmatrix} ,
\end{equation} 
where the energy difference of the two-level system is given by $\hbar
\delta = -g \mu {\cal B}_z$, where ${\cal B}_z$ is the static magnetic
field, and the Rabi frequency $\Omega$ is proportional to the
electromagnetic field strength that oscillates at frequency $\omega$.
Denoting the stochastic magnetic field as ${\bf B}_{\mathrm{st}}(t)$,
the stochastic Hamiltonian takes the form,
\begin{equation} \label{Eq:H_stochastic}
    H_{\mathrm{st}}(t) = -\frac{g \mu}{2} {\bf B}_{\mathrm{st}}
    (t) \cdot {\boldsymbol \sigma} ,
    \end{equation} 
where ${\boldsymbol \sigma} = (\sigma_x, \sigma_y, \sigma_z)$ is the
vector of Pauli spin matrices.  The average of ${\bf B}(t)$ over the
stochastic fluctuations is taken to vanish, and the field correlation
function depends upon the type of noise \cite{vanKampenBook},
\begin{equation}  \label{Eq:mean_var}
    \overline{{\bf B}_{\mathrm{st}}(t)} = 0 , \quad
    \overline{B_{i,\mathrm{st}}(t) B_{j,\mathrm{st}} (t')} = 
    k_{ij}(t-t') , \ \ i,j = x, y, z ,
\end{equation}
where $\overline{(\ldots)}$ denotes the stochastic average, and
$k_{ij}(t-t')$ is the stochastic field correlation function.  The full
Hamiltonian for the two-level system is $\tilde H(t) = H(t) +
H_{\mathrm{st}}(t)$.  There is a considerable literature on the use of
the RWA in such problems \cite{Fleming_12b, Accardi_91, Villaeys_91, 
Grifoni_95, Stenius_96, Prants_99, Cao_03, Yu_04, Ishizaki_05, Ishizaki_06,
Schmidt_11, Doherty_12, Fleming_10, Fleming_12, Fleming_12b, Desposito_92}, 
and we shall explore the stochastic dynamics with and without making the 
RWA.

Specifically, here we explore the stochastic approach, and, for
Gaussian white noise, the master equation approach, to the problem.
We explicitly consider white Gaussian noise (Wiener processes) and
colored Gaussian noise (Ornstein--Uhlenbeck processes).  The outline
of the paper is as follows.  In Sec.~\ref{Sec:DOF}, in order to
set out the notation used in this paper and to compare
with the stochastic dynamics in the coming sections, we present
results for the dynamics of the two-level system in an oscillating
field without any stochasticity present, both without and with making
the RWA. We discuss the stochastic dynamics in
Sec.~\ref{Sec:Stoch_Dynamics}, first treating dephasing due to white
noise in the transverse magnetic field ($b_z$) in
Sec.~\ref{SubSec:Dephasing}, then isotropic white noise in
Sec.~\ref{SubSec:Iso_w_n}.  In Sec.~\ref{Sec:Master_Eqs} we present
the master (Liouville--von Neumann) equation results for Gaussian
white noise.  We find that the RWA transformation to the rotating
frame does not commute with the stochastic Hamiltonian when the noise
has components along all coordinate directions.  This has the
potential for affecting the results obtained using the RWA in both
stochastic dynamics and master equation dynamics, but we find that for
Gaussian white noise, the effect is negligible.  We find an analytic
solution to the density matrix dynamics for Gaussian white noise.
Section~\ref{Sec:OU} considers Ornstein--Uhlenbeck noise, and for this
case isotropic noise of this kind, the non-commutation of the RWA
transformation with the stochastic Hamiltonian need not be negligible.
Finally, a summary and conclusion is presented in
Sec.~\ref{Sec:Summary}.  This section also contains an explicit
example of a rather well-studied physical system, nitrogen vacancy
centers in diamond driven by an electromagnetic field, in which the
field induces transitions between levels that are subject to a noisy 
environment. The reader desiring motivation for the model used here
prior to learning the details of the model is encouraged to first
read the last paragraph of Sec.~\ref{Sec:Summary}.

\section{Dynamics in an Oscillating field and the Rotating Wave 
Approximation}  \label{Sec:DOF}

The time-dependent Schr\"odinger equation for our two-level system is,
$i \hbar \dot {\psi} = \tilde H(t) \psi$, where $\psi(t) = 
\binom{\psi_b(t)} {\psi_a(t)}$ is the two-component solution and $\tilde 
H(t)$ is the time-dependent Hamiltonian given by the sum of (\ref{Eq:H}) 
and (\ref{Eq:H_stochastic}).  In this section, for the sake of comparison
with the stochastic dynamics to be presented in
Secs.~\ref{Sec:Stoch_Dynamics} and \ref{Sec:OU}, and the master
equation results in Sec.~\ref{Sec:Master_Eqs}, we discuss the
treatment of the problem without a stochastic Hamiltonian, both
without and with making the RWA (i.e., transforming to the rotating
frame wherein the Hamiltonian ${\cal H}_{\mathrm{RWA}}$ is
time-independent).  The time-dependent Schr\"odinger equation will be 
solved with the initial condition $\psi(0) \equiv \binom{\psi_b(0)}
{\psi_a(0)} =\binom{1}{0}$ at time $t = 0$.  
The probabilities for being in states $b$ and $a$ at time $t$
are given by $P_b(t) = |\psi_b(t)|^2$ and $P_a(t) = |\psi_a(t)|^2$.
The inset in Figs.~\ref{Fig_stochastic_RWA_sz_d_1_W_0.2}(c) and (d)
show the calculated probability $P_b(t) = |\psi_b(t)|^2$ versus time
for the on-resonance case, $\delta/\omega = 1$, and Rabi frequency
$\Omega/\omega = 0.2$, without and with making the RWA. For any
detuning $\delta$, the probabilities oscillate (Rabi-flop) with
generalized Rabi frequency $\Omega_g = \sqrt{\Omega^2 + \Delta^2}$,
where $\Delta = \omega - \delta$ is the detuning from resonance.
Moreover, without making the RWA, there is a fast oscillation at
frequency $\omega + \delta$, which is clearly evident, and there is
also a Bloch--Siegert shift of the resonance frequency by $\delta
\omega_{\mathrm{BS}} = \Omega^2/(4 \omega)$ \cite{Bloch_Siegert_40}.
The insets show that, for the on-resonance case, $\Delta = 0$, aside
from the additional oscillations due to the high frequency components
and the small Bloch--Siegert shift (which is barely visible here,
since $\omega_{\mathrm{BS}} = 0.01$), the nature of the RWA dynamics
is rather similar to that obtained without making the RWA.


If $\delta \approx \omega$, one often makes the rotating wave
approximation (RWA), wherein one transforms to a rotating frame
wherein the Hamiltonian, after neglecting a quickly oscillating
component, is approximately time-independent.  Letting the
transformation to the rotating frame, ${\cal U}(t)$, be such that
\cite{Band_Avishai}
\begin{equation} \label{Eq:RWA}
    \psi(t) = {\cal U}(t) \varphi(t), \quad {\cal U}(t) = 
    \begin{pmatrix} e^{-i\delta_b t} & 0 \\
    0 & e^{-i\delta_a t} \end{pmatrix} ,
\end{equation}
taking $\delta_a = - \delta/2$ and $\delta_b = \omega + \delta_a$,
and noting that
\begin{equation} \label{Eq:RWA_Sch}
    i \hbar \frac{\partial \varphi(t) }{\partial t} = [{\cal
    U}^\dagger(t) {\cal H} {\cal U}(t)-i \hbar \, {\cal U}^\dagger(t)
    \dot {\cal U}(t)] \varphi(t) ,
\end{equation}
and dropping quickly oscillatory terms, yields the following
Schr\"odinger equation for the spinor $\varphi(t)$:
\begin{equation}  \label{Eq:Sch_RWA}
    i \frac{d}{dt} 
    \left(\! \! \begin{array}{c}
	\varphi_b \\ \varphi_a \end{array} \! \! \right) = 
    \left( \! \! \begin{array}{cc}
	-\Delta & i \frac{\Omega}{2}  \\
	-i \frac{\Omega}{2} & 0 \end{array} \! \! \right)
    \left(\! \! \begin{array}{c} \varphi_b \\ \varphi_a \end{array} 
    \! \! \right) ~.
\end{equation}
Applying a further transformation, $\varphi_a \to -i \varphi_a$, the
full RWA transformation becomes
\begin{equation}  \label{Eq:RWA_trans}
    {\cal U}(t) = e^{i\delta \, t/2} \begin{pmatrix} e^{- i \omega t}
    & 0 \\ 0 & -i \end{pmatrix} .
\end{equation}
This last transformation turns the complex Hermitian time-independent
Hamiltonian matrix on the RHS of (\ref{Eq:Sch_RWA}) into a real
symmetric time-independent Hamiltonian, and the Schr\"odinger 
equation becomes \cite{Band_Avishai},
\begin{equation}  \label{Eq:Sch_RWA'}
    i \frac{d}{dt} 
    \left(\! \! \begin{array}{c}
	\varphi_b \\ \varphi_a \end{array} \! \! \right) = 
    \left( \! \! \begin{array}{cc}
	-\Delta & \frac{\Omega}{2}  \\
	\frac{\Omega}{2} & 0 \end{array} \! \! \right)
    \left(\! \! \begin{array}{c} \varphi_b \\ \varphi_a \end{array} 
    \! \! \right) ~.
\end{equation}
Hence, the (constant) RWA Hamiltonian matrix is  
$H_{\mathrm{RWA}} \equiv \hbar \left(\!  \!  \begin{array}{cc} -\Delta &
\Omega/2 \\ \Omega/2 & 0 \end{array} \!  \!  \right)$.  The criteria
for the validity of the RWA are $|\Delta| \ll \omega$ and $\Omega < 
\omega$.

In the remainder of this paper, we use dimensionless quantities; we 
set $\hbar = 1$, take time to be measured in units of $1/\omega$, and the 
frequencies $\delta$ and $\Omega$ to be in units of $\omega$ (i.e., we 
take $\omega = 1$).  The dimensionless system Hamiltonian is given by 
\begin{equation} \label{Eq:H'}
    {\cal H}(t) = \begin{pmatrix} \frac{\delta}{2} & \Omega \sin
     t \\ \Omega \sin t & - \frac{\delta}{2} \end{pmatrix} , 
\end{equation}
the dimensionless stochastic Hamiltonian is
\begin{equation} \label{Eq:H_stochastic'}
    {\cal H}_{\mathrm{st}}(t) = {\bf b}(t) \cdot
    {\boldsymbol \sigma} = \begin{pmatrix} b_z(t) & b_x(t) - i b_y(t)
    \\ b_x(t) + i b_y(t) & - b_z(t) \end{pmatrix},
\end{equation}
where ${\bf b}(t)$ is the dimensionless stochastic magnetic field, and
the full dimensionless Hamiltonian is $\tilde {\cal H}(t) = {\cal
H}(t) + {\cal H}_{\mathrm{st}}(t)$.  The RWA Hamiltonian is ${\cal
H}_{\mathrm{RWA}} = \left(\!  \!  \begin{array}{cc} -\Delta & \Omega/2
\\ \Omega/2 & 0 \end{array} \!  \!  \right)$, where the dimensionless
detuning is $\Delta = \omega - \delta$ (i.e., $\Delta = 1 -
\delta/\omega$), and the dimensionless Rabi frequency is $\Omega$
(i.e., $\Omega/\omega$).  An important parameter regarding the
stochastic magnetic field is the noise correlation time
$\tau_{\mathrm{corr}}$, which is determined by the nature of the
noise.  For Gaussian white noise, $\tau_{\mathrm{corr}}$ is
infinitesimal, but for an OU process (colored Gaussian noise) $\tau_{\mathrm{corr}}$
is an important parameter that characterizes the noise.  We shall see
in Secs.~\ref{SubSec:Iso_w_n} and \label{SubSec:OU} that an important
dimensionless parameter that characterizes the response of the system
to the noise is $\omega \tau_{\mathrm{corr}}$ (in dimensionless time
units, $\tau_{\mathrm{corr}}$).

\section{Stochastic Dynamics} \label{Sec:Stoch_Dynamics}

There are a number of ways of modeling stochastic processes, including
a master equation method \cite{master_eq}, a Monte Carlo wave function
method \cite{Molmer_93}, and a stochastic differential equations
method \cite{vanKampenBook, Gardiner, Kloeden, Kloeden_03}.  In this
section, we use stochastic differential equations.

If the characteristic timescale of the fluctuations is much shorter
than the timescale of free evolution of the system, the noise
correlations can be well approximated by a Dirac $\delta$ function to
obtain the {\em Gaussian white noise limit} wherein the dimensionless
correlation functions $\kappa_{ij}(\tau)$ (related to the dimensional
correlation functions appearing in Eq.~(\ref{Eq:mean_var}) are
proportional to Dirac $\delta$ functions.  If the noise in
the different components of the magnetic field is uncorrelated,
\begin{equation}  \label{Eq:white-noise}
    \overline{b_i(t) b_j(t')} = \kappa_{ij}(t-t') = w_{0,i}^2 \,
    \delta_{ij} \, \delta(t-t') .
\end{equation}
The quantity $w_{0,i}$ is the dimensionless volatility of the $i$th
component of the dimensionless stochastic field ${\bf b}(t)$.

A Wiener process $w(t)$ is the integral over time of white noise
$\xi(t)$, i.e., $\xi(t) = d w(t)/dt$, with $\overline{\xi(t)} = 0$ and
$\overline{\xi(t) \xi(t')} = w_{0}^2 \delta(t-t')$ [compare with
Eq.~(\ref{Eq:white-noise})].  Thus, the stochastic magnetic field
components are taken to be the time-derivative of a Wiener process.
The SLSDE for a quantum system coupled to
a Wiener stochastic process $w(t)$ via operator ${\cal V}$ is given by
\cite{vanKampenBook},
\begin{equation}  \label{Schr_Langevin}
    {\dot \psi} = -i {\cal H} \psi + w_{0} \, \xi(t) {\cal V}
    \psi - \frac{w_{0}^2}{2} {\cal V}^{\dag} {\cal V} \psi .
\end{equation}
The $w_{0}^2$ term in Eq.~(\ref{Schr_Langevin}) insures unitarity if 
${\cal V}$ is a Hermitian operator \cite{vanKampenBook}.  Equation
(\ref{Schr_Langevin}) can be easily generalized to include {\em sets} of
operators ${\cal V}_i$, stochastic processes $w_i(t)$, and
volatilities $w_{0,i}$, to obtain the {\em general SLSDE},
\begin{equation}  \label{Schr_Langevin_gen}
    {\dot \psi} = - i {\cal H} \psi + \sum_i\left( w_{0,i} \xi_i(t)
    {\cal V}_i \psi - \frac{w_{0,i}^2}{2} {\cal V}_i^{\dag} {\cal V}_i
    \psi \right) .
\end{equation}
Equation (\ref{Schr_Langevin}) is equivalent to a Markovian quantum
master equation with a Lindblad operator ${\cal V}$, and the more
general Eq.~(\ref{Schr_Langevin_gen}) is equivalent to the Markovian 
quantum master equation
\begin{equation}  \label{Eq:master}
    {\dot \rho} = -i [{\cal H}, \rho(t)] + \frac{1}{2} \sum_j w_{0,j}^2
    \left(2{\cal V}_j \rho(t) {\cal V}^{\dag}_j - \rho(t) {\cal
    V}^{\dag}_j {\cal V}_j - {\cal V}^{\dag}_j {\cal V}_j \rho(t)
    \right) ,
\end{equation}
with the set of Lindblad operators ${\cal V}_i$ \cite{master_eq, 
vanKampenBook}.

For example, for the dephasing case to be studied in
Sec.~\ref{SubSec:Dephasing}, the Lindblad operator is taken to be
${\cal V} = \sigma_z$, and the wave function $\psi$ is a two component
spinor. In stochastic process notation \cite{vanKampenBook, Gardiner, 
Kloeden, Kloeden_03}, Eq.~(\ref{Schr_Langevin}), takes the form
\begin{subequations}  \label{Eq:TLS_stoch_dephas}
\begin{equation}  \label{Eq:TLS_stoch_dephas1}
    d \psi_{b}(t) = \left\{[ -i({\cal H}_{bb} \psi_{b}(t) + {\cal
    H}_{ba} \psi_{a}(t)) - \frac{w_0^2}{2} \psi_b(t)] \, dt + w_0
    \psi_b(t) \, dw \right\},
\end{equation}
\begin{equation}  \label{Eq:TLS_stoch_dephas2}
    d \psi_{a}(t) = \left\{[ -i( {\cal H}_{ab} \psi_{b}(t) + {\cal
    H}_{aa} \psi_{a}(t)) - \frac{w_0^2}{2} \psi_a(t)] \, dt - w_0
    \psi_a(t) \, dw \right\} .
\end{equation}
\end{subequations}
For any specific realization of the stochastic process, these
equations are solved to yield the two component spinor $\binom
{\psi_b(t)} {\psi_a(t)}$ (which is itself a stochastic variable).  The
(survival) probability to be in state $b$ at time $t$ is
$P(t)=|\psi_b(t)|^2$.  The distinction, as compared with the
deterministic case ($w_0=0$), is that now $P(t)$ is a random function
with distribution ${\cal D}[P(t)]$.  Equations
(\ref{Eq:TLS_stoch_dephas}) are easily generalizable to white noise in
all three components of the magnetic field; the Lindblad operators
appearing in (\ref{Schr_Langevin_gen}) are then ${\cal V}_i =
\sigma_i$ and $w_{0,i}$ are the volatilities for $b_i(t)$.  For the
isotropic case (treated in Sec.~\ref{SubSec:Iso_w_n}), the numerical
values of $w_{0,i}$ are equal.

\subsection{Dephasing due to Transverse White Noise}
\label{SubSec:Dephasing}

Dephasing of a system occurs due to interaction between the system and
its environment which scrambles the phases of the wave function of the
system without {\em directly} affecting probabilities.  One of the
methods for treating dephasing of a quantum system is to model the
interaction with the environment in terms of a time-dependent random
noise.  Such an approach enables the calculation of not only the averaged
survival probability, $\overline{P(t)}$, but also its standard
deviation, $\Delta P(t) = \left(\overline{P(t)^2} - (\overline{P(t)})^2 
\right)^{1/2}$, and its statistical distribution function 
${\cal D}[P(t)]$.  When the fluctuating magnetic field has a 
non-vanishing component only along $z$, Eq.~(\ref{Eq:H_stochastic}) 
reduces to
\begin{equation} \label{Eq:H_stochastic_z}
    {\cal H}_{\mathrm{st}}(t) = \xi(t) \sigma_z ,
\end{equation}
i.e., ${\bf b}(t) = \xi(t) \hat {\bf z}$ where $\xi(t)$ can be taken
as white noise, which has an infinitesimal correlation time, if the
correlation time of the bath, $\tau_{\mathrm{corr}}$, is very fast in
comparison to the timescales of the spin system, so, to a good
approximation,
\begin{equation}  \label{Eq:white_noise}
    \overline{\xi(t)} = 0, \quad \overline{\xi(t) \xi(t')} =
    w_0^2 \delta(t-t') .
\end{equation}
As discussed earlier in connenction with Eq.~(\ref{Eq:white-noise}),
the white noise $\xi(t)$ can be written as the time derivative of a
Wiener process $w(t)$, $\xi(t) = d w(t)/dt$, or more formally, the
Wiener process $w(t)$ is the integral of the white noise.  The
stochastic Hamiltonian in Eq.~(\ref{Eq:H_stochastic_z}) gives rise to
dephasing of the wave function of the two-level system.  In the case
of dephasing due to collisions with particles, each collision can have
a random duration and a random strength, and in the case of
interactions with an environment, the many degrees of freedom of the
environment can randomly affect the phase of the wave functions
$\psi_a(t)$ and $\psi_b(t)$.  This results in a time-dependent
uncertainty $\delta \varphi_j(t)$ in the phase of the wave function
component $\psi_j(t)$.  At a time $t=\tau$ for which $\delta
\varphi_j(\tau) = 2 \pi$, interference is completely lost.  The
volatility (the stochastic field strength) $w_0$ appearing in
Eq.~(\ref{Eq:white_noise}) is inversely proportional to the
correlation time $\tau_\phi \equiv \tau_{\mathrm{corr}}$ of the bath.
Incorporation of dephasing in two-level system dynamics has been
extensively studied \cite{Rammer, Efrat1, Efrat2, Pokrovsky, Avron}.

\begin{figure}
\centering\subfigure[]{\includegraphics[width=0.45\textwidth]
{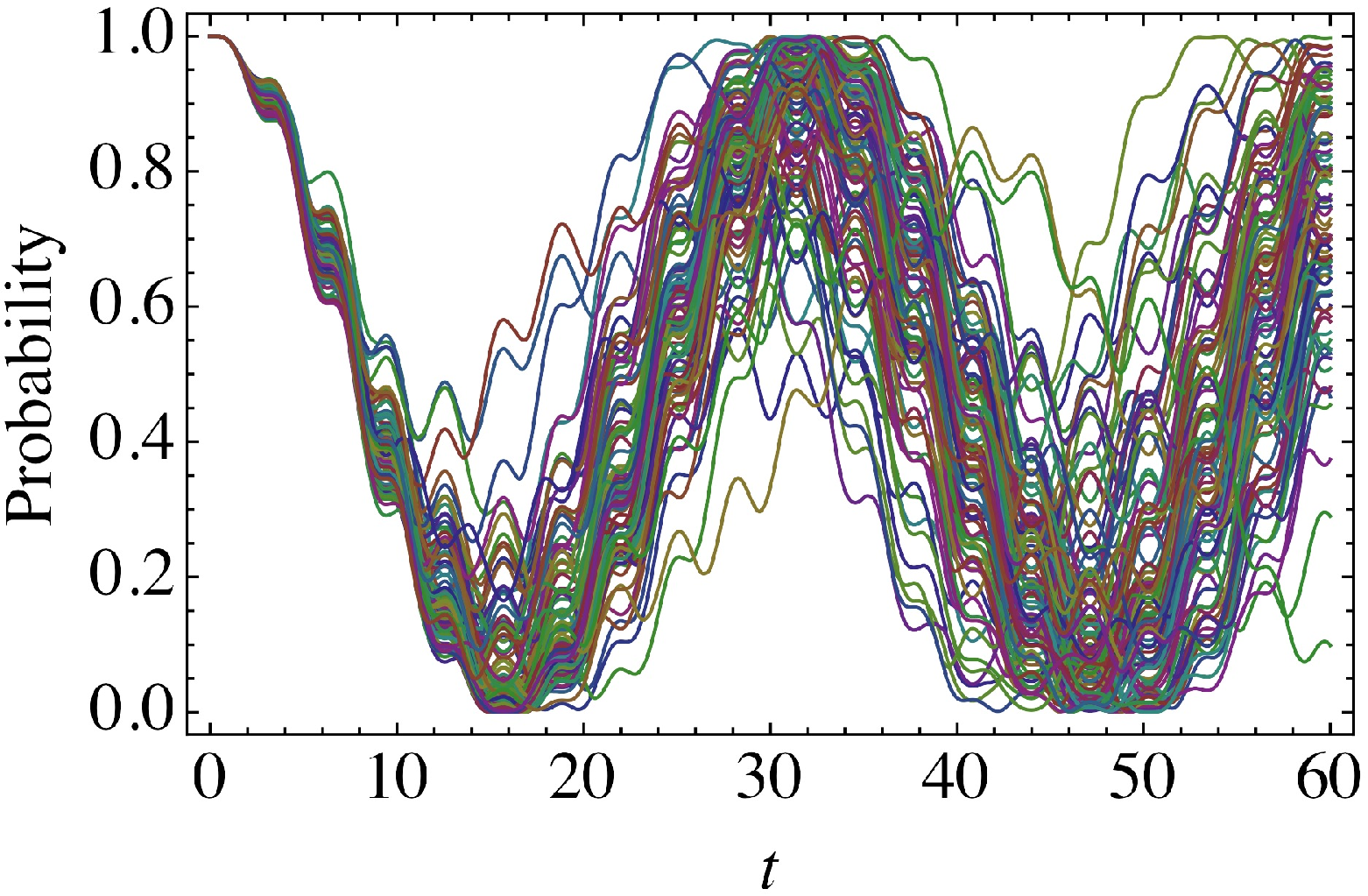}}
\centering\subfigure[]{\includegraphics[width=0.45\textwidth]
{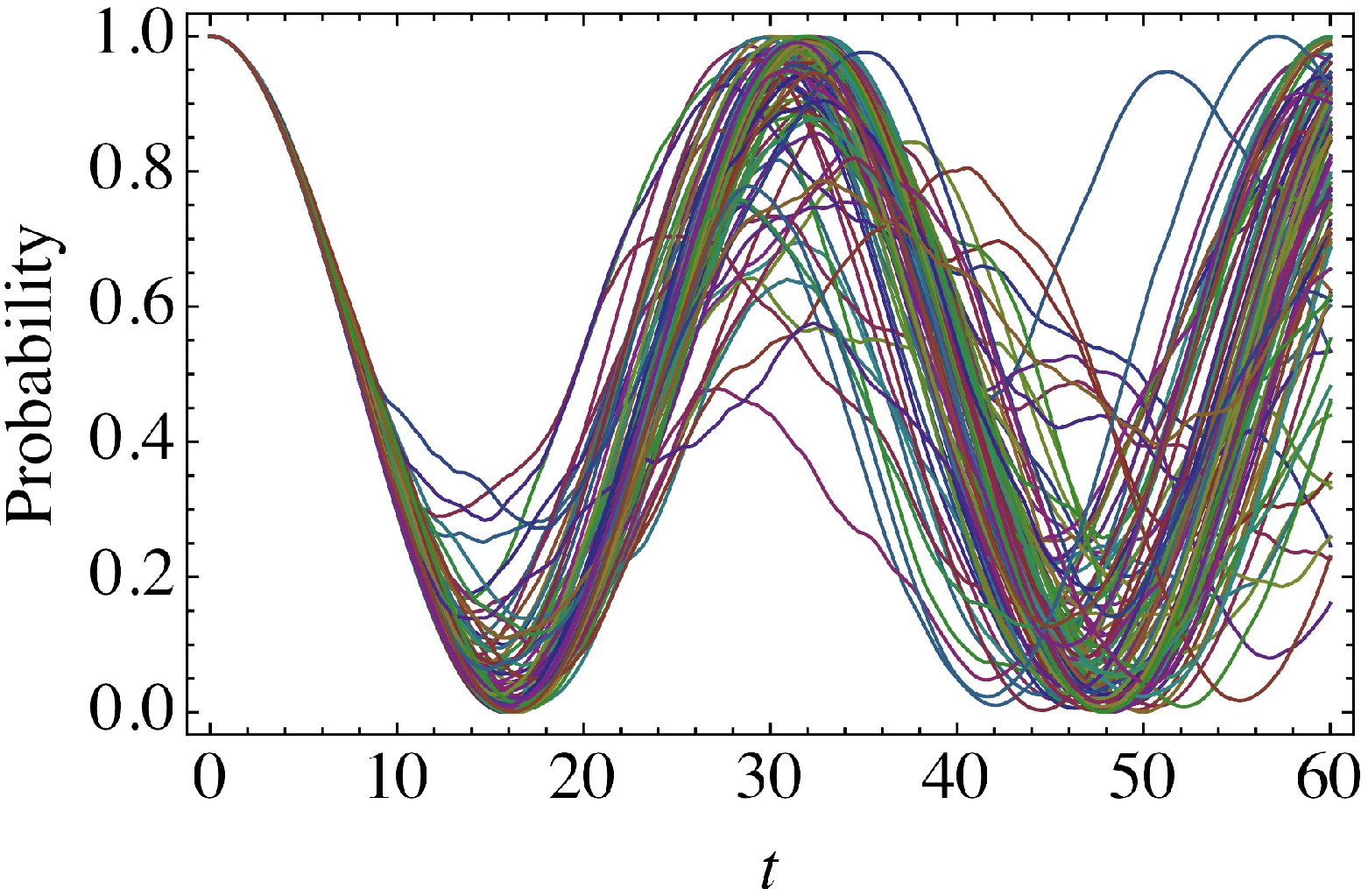}}
\centering\subfigure[]{\includegraphics[width=0.45\textwidth]
{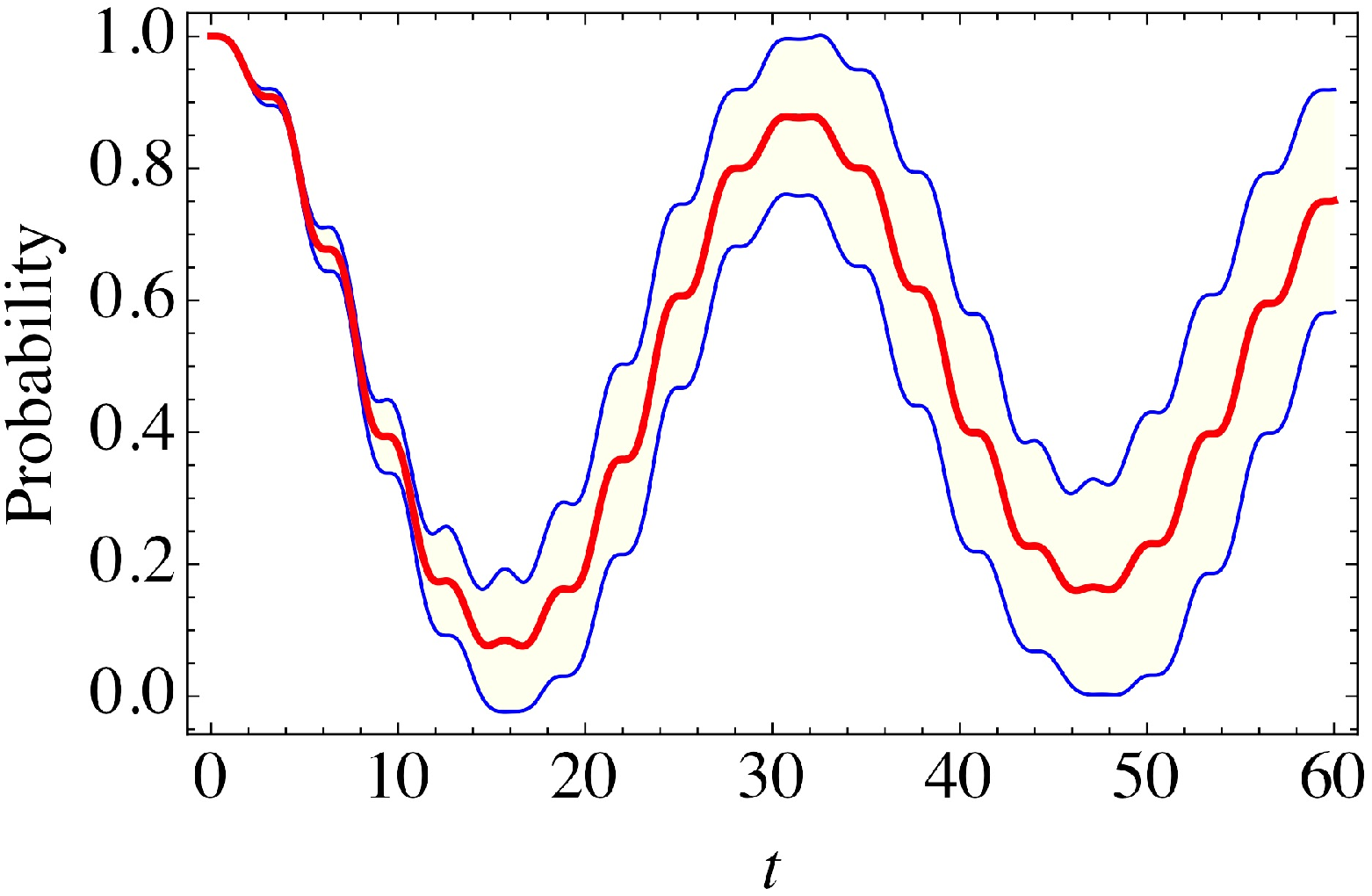}}
\centering\subfigure[]{\includegraphics[width=0.45\textwidth]
{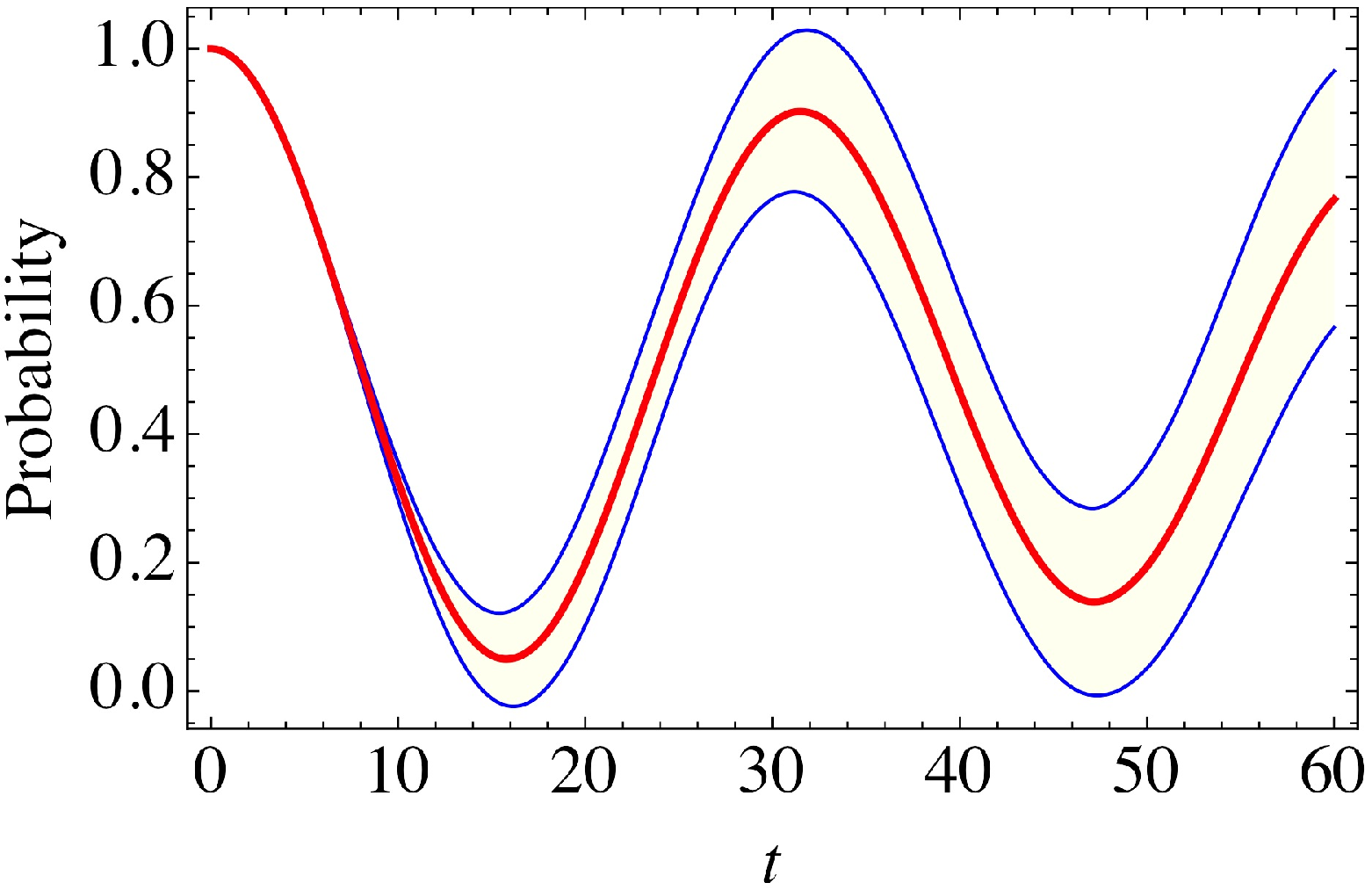}}
\caption{(Color online) Dephasing of on-resonance transitions due to a
stochastic field $b_z(t)$.  (a) 100 stochastic realizations of the
probability $P_b(t) = |\psi_b(t)|^2$ versus time for the on-resonance
case, for $\delta=1.0$, $\omega=1.0$ and $\Omega=0.2$ in the presence
of a stochastic magnetic field along the $z$ direction with volatility
$w_0 = 0.1$.  (b) 100 stochastic realizations of the rotating wave
approximation for the probability $P_b(t)$ versus time for the
on-resonance case, for $\Delta=0$, $\omega=1.0$ and $\Omega=0.2$ in
the presence of a stochastic magnetic field along the $z$ direction
with volatility $w_0 = 0.1$.  (c) Average probability
$\overline{P_b(t)} = \overline{\psi_b^*(t) \psi_b(t)}$ and the average
plus and minus standard deviation of the probability versus time for
$\delta=1.0$, $\omega=1.0$ and $\Omega=0.2$ and a stochastic field in
the $z$ direction with volatility $w_0 = 0.1$.  (d) Average rotating
wave approximation probability $\overline{P_b(t)}$ and the average
plus and minus standard deviation of the probability versus time for
$\Delta=0$, $\omega=1.0$ and$\Omega=0.2$ and a stochastic field in the
$z$ direction with volatility $w_0 = 0.1$.  For comparison, the insets 
in (c) and (d) show the results without any noise.}
\label{Fig_stochastic_RWA_sz_d_1_W_0.2}
\end{figure}

\begin{figure}
\centering\subfigure[]{\includegraphics[width=0.45\textwidth]
{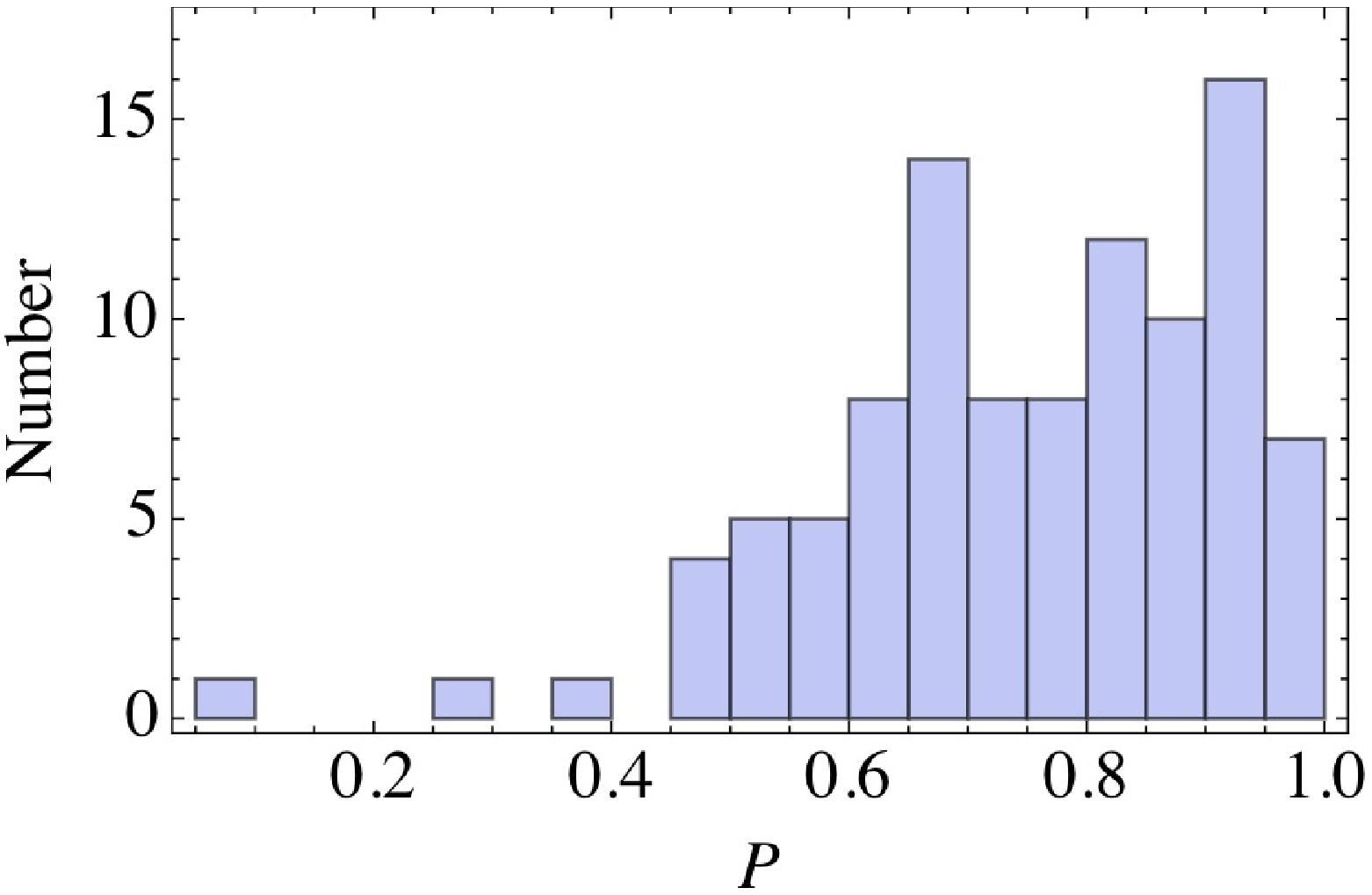}}
\centering\subfigure[]{\includegraphics[width=0.45\textwidth]
{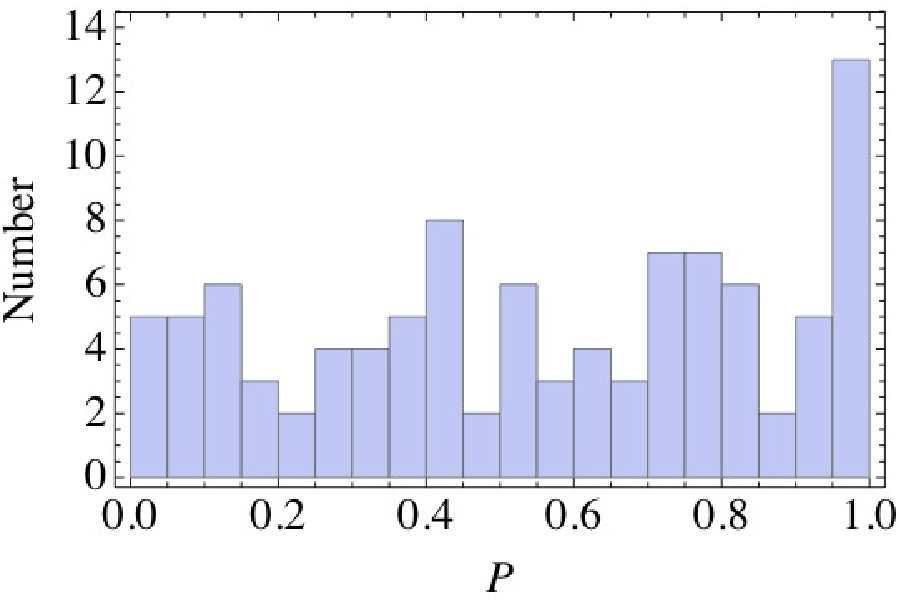}} 
\caption{(Color online) (a) Histogram with 100 paths (realizations) 
of the probability $P_b(t)$ in the presence of a stochastic magnetic 
field along the $z$ direction at the final computed time, $t = 60$ 
shown in Figs.~\ref{Fig_stochastic_RWA_sz_d_1_W_0.2}(a) and (c).  
(b) Histogram  with 100 paths (realizations) of 
the probability $P_b(t)$ in the presence of an isotropic stochastic 
magnetic field at the final computed time, $t = 60$ to be shown in
Fig.~\ref{Fig_stochastic_nonRWA_d_1_W_0.2}(a).}
\label{Fig_nonRWA_dist}
\end{figure}

Our stochastic calculations were carried out using the Mathematica 9.0
built-in command {\em ItoProcess} \cite{ItoProcess}.
Figure~\ref{Fig_stochastic_RWA_sz_d_1_W_0.2} shows the results for a
stochastic magnetic field in the $z$ direction that corresponds to
white noise with volatility $w_0 = 0.1$.  Specifically,
Figs.~\ref{Fig_stochastic_RWA_sz_d_1_W_0.2}(a) and (b) show 100
stochastic realizations of the probability $P_b(t) = |\psi_b(t)|^2$
versus time for the on-resonance case, $\delta=1.0$, $\omega=1.0$ and
$\Omega=0.2$, computed without and with making the RWA.
Figures~\ref{Fig_stochastic_RWA_sz_d_1_W_0.2}(c) and (d) show the mean
probabilities and the standard deviations for these cases.  For very
large time, the oscillations in the probabilities die out and the
probabilities go to 1/2.  Figure~\ref{Fig_nonRWA_dist}(a) shows the
histogram of the probability $P_b(T)$ distribution, ${\cal D}[P_b(T)]$,
at the final computed time, $T = 60$, for the case shown in
Fig.~\ref{Fig_stochastic_RWA_sz_d_1_W_0.2}(a).

\subsection{Decoherence due to Isotropic White Noise} \label{SubSec:Iso_w_n}

\begin{figure}
\centering\subfigure[]{\includegraphics[width=0.3\textwidth]
{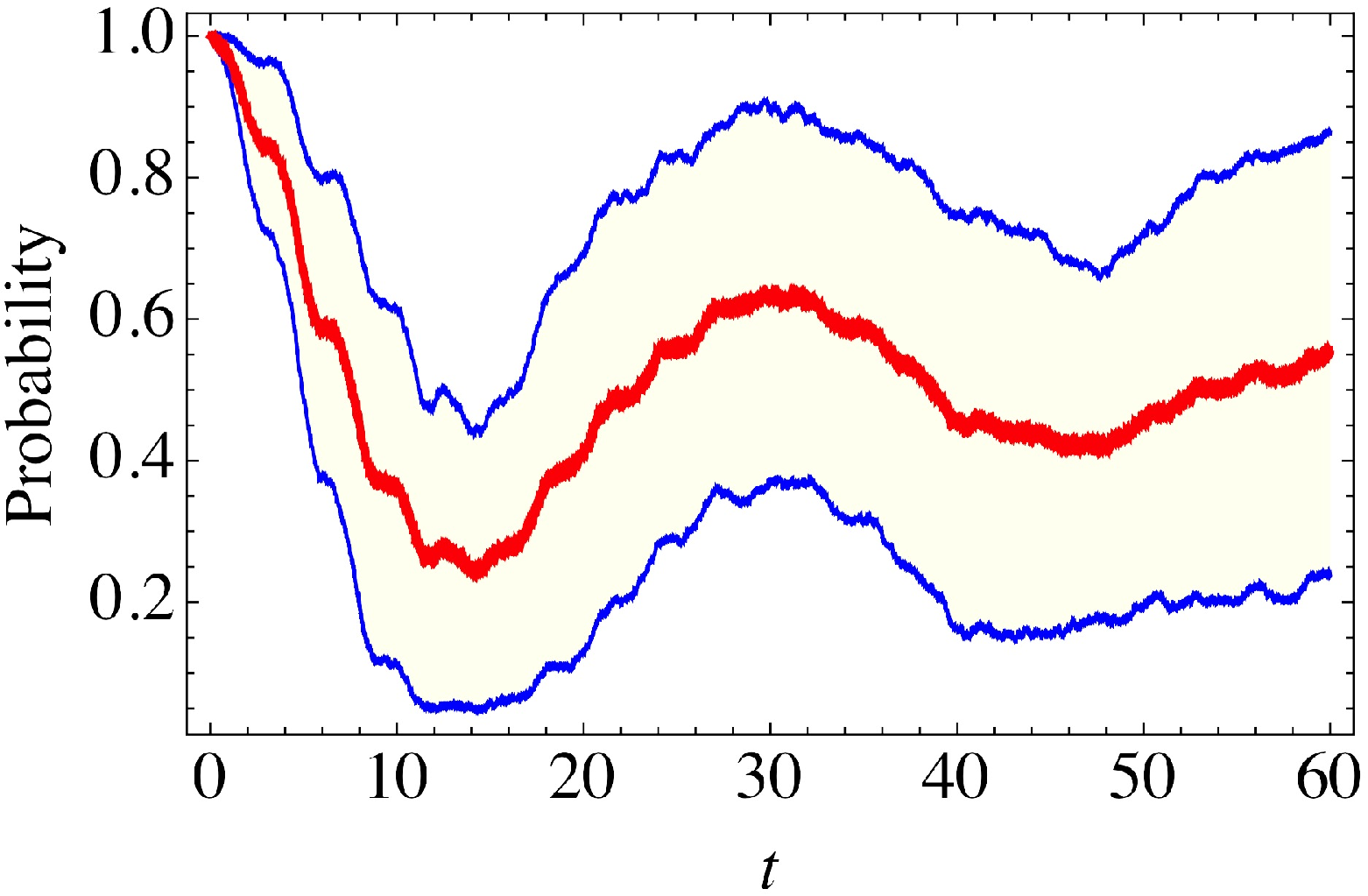}}
\centering\subfigure[]{\includegraphics[width=0.3\textwidth]
{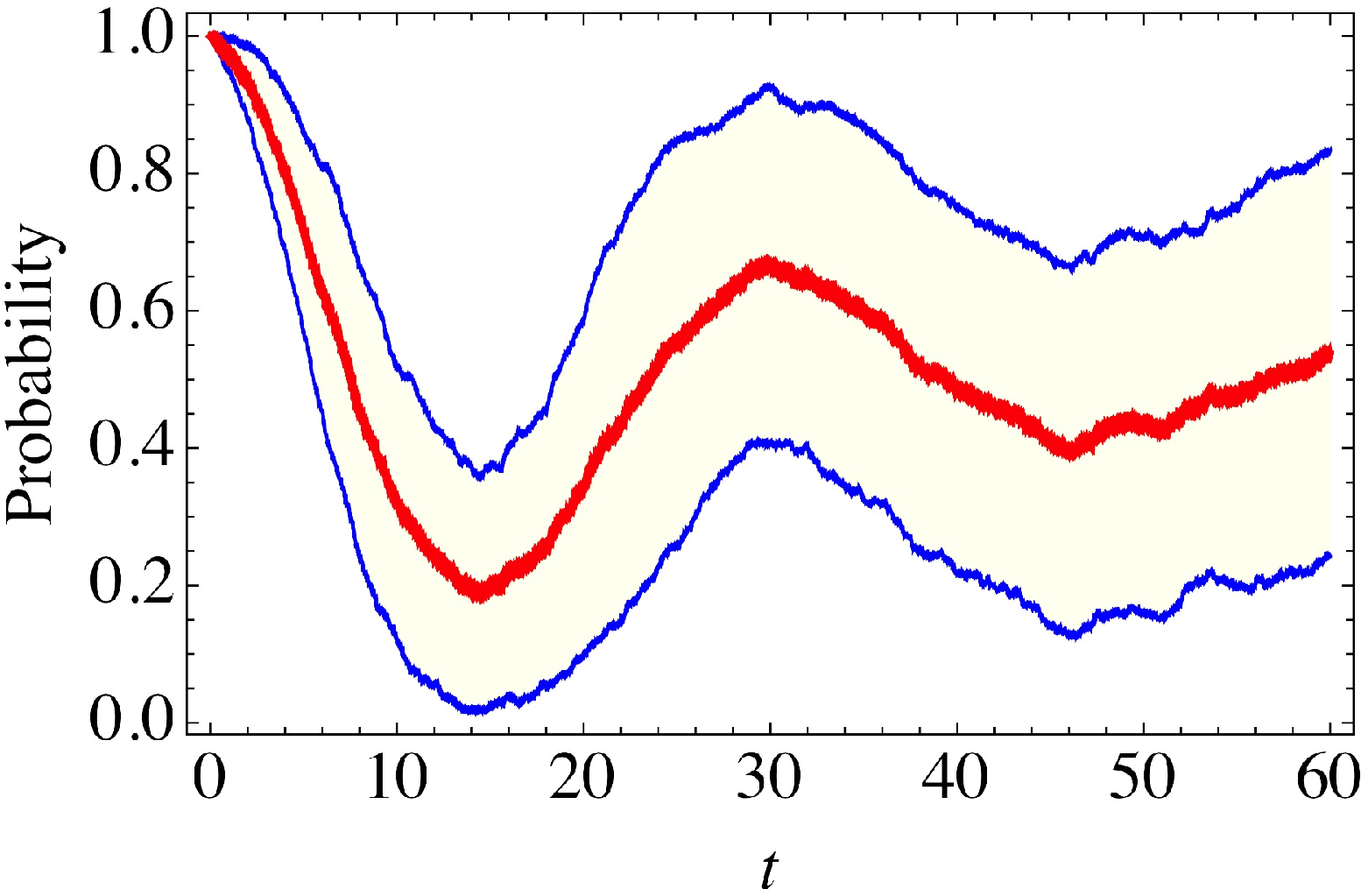}}
\centering\subfigure[]{\includegraphics[width=0.3\textwidth]
{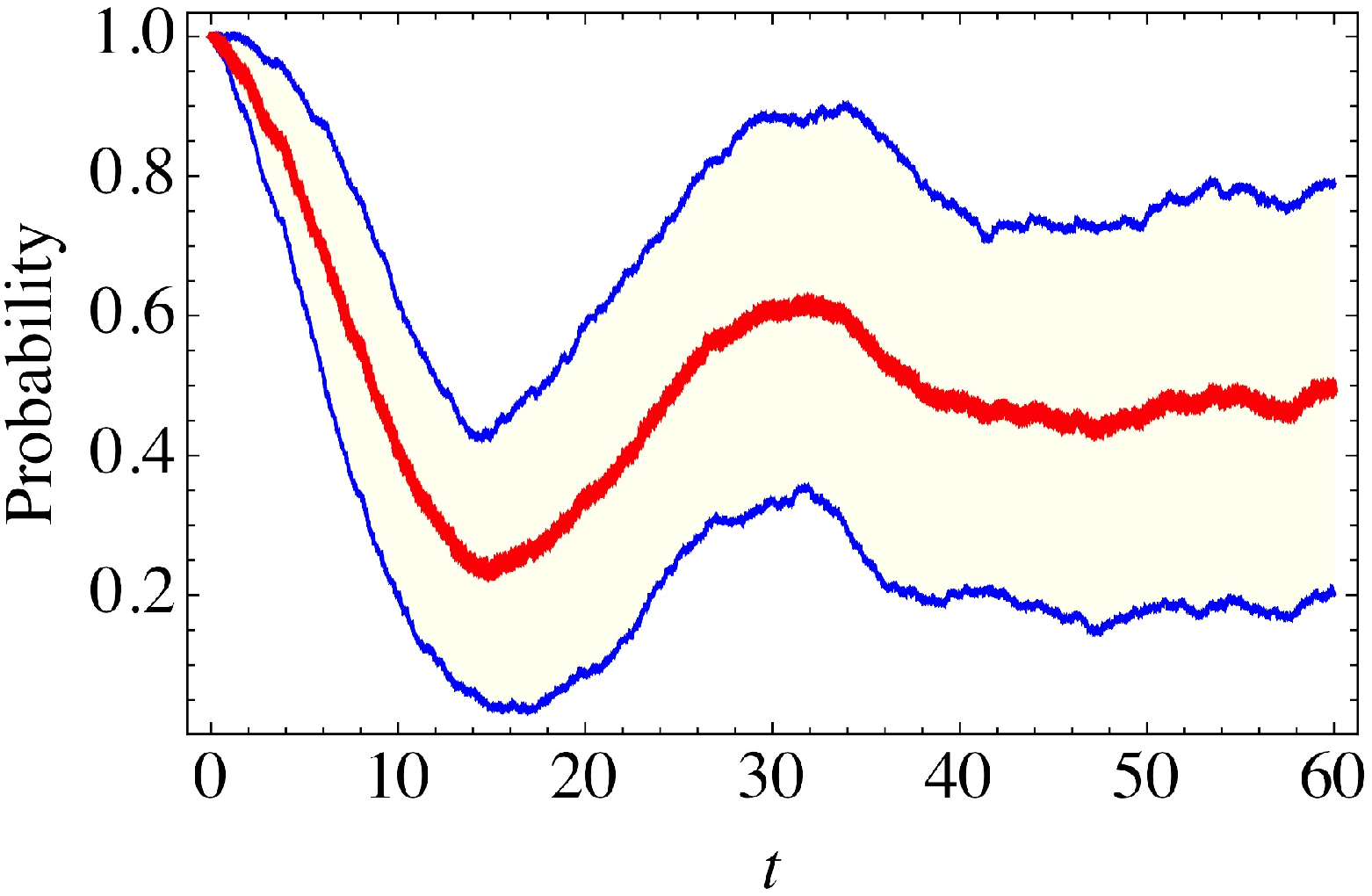}} 
\caption{(Color online) Decoherence and dephasing of on-resonance
transitions with isotropic white noise.  (a) Average probability
$\overline{P_b(t)} = \overline{\psi_b^*(t) \psi_b(t)}$ and the
standard deviation of the probability versus time for $\delta=1.0$,
$\omega=1.0$, $\Omega=0.2$ and volatilities $w_{0,1} = w_{0,2} =
w_{0,3} = 0.1$.  (b) Same as (a) (i.e., $\Delta = 0$), except
calculated using the RWA. (c) Same as (b), except that the
non-commutativity of the RWA transformation and the transverse
stochastic Hamiltonian not properly accounted for.}
\label{Fig_stochastic_nonRWA_d_1_W_0.2}
\end{figure}

Figure~\ref{Fig_stochastic_nonRWA_d_1_W_0.2}(a) shows the average
probability $\overline{P_b(t)} = \overline{\psi_b^*(t) \psi_b(t)}$,
and the average plus and minus the standard deviation of the
probability calculated using Eq.~(\ref{Schr_Langevin_gen}) in the form
\begin{equation}  \label{Schr_Langevin_gen'}
    {\dot \psi} = \left(-i \, {\cal H} - \frac{3 w_0^2}{2} {\bf 1} \right) \psi
     + w_0 \sum_i \xi_i(t) \sigma_i \, \psi ,
\end{equation}
where the white noise $\xi_i(t)$ satisfy $\overline{\xi_i(t)} = 0$ and
$\overline{\xi_i(t) \xi_j(t')} = \delta_{ij} \, \delta(t-t')$, with
the volatilities $w_{i,0} \equiv w_0 = 0.1$ for $i = 1, 2, 3$.
Figure~\ref{Fig_nonRWA_dist}(b) shows the histogram of the probability
$P_b(t)$ at the final computed time, $t = 60$, for the case shown in
Fig.~\ref{Fig_stochastic_nonRWA_d_1_W_0.2}(a).

\begin{figure}
\centering\subfigure[]{\includegraphics[width=0.3\textwidth]
{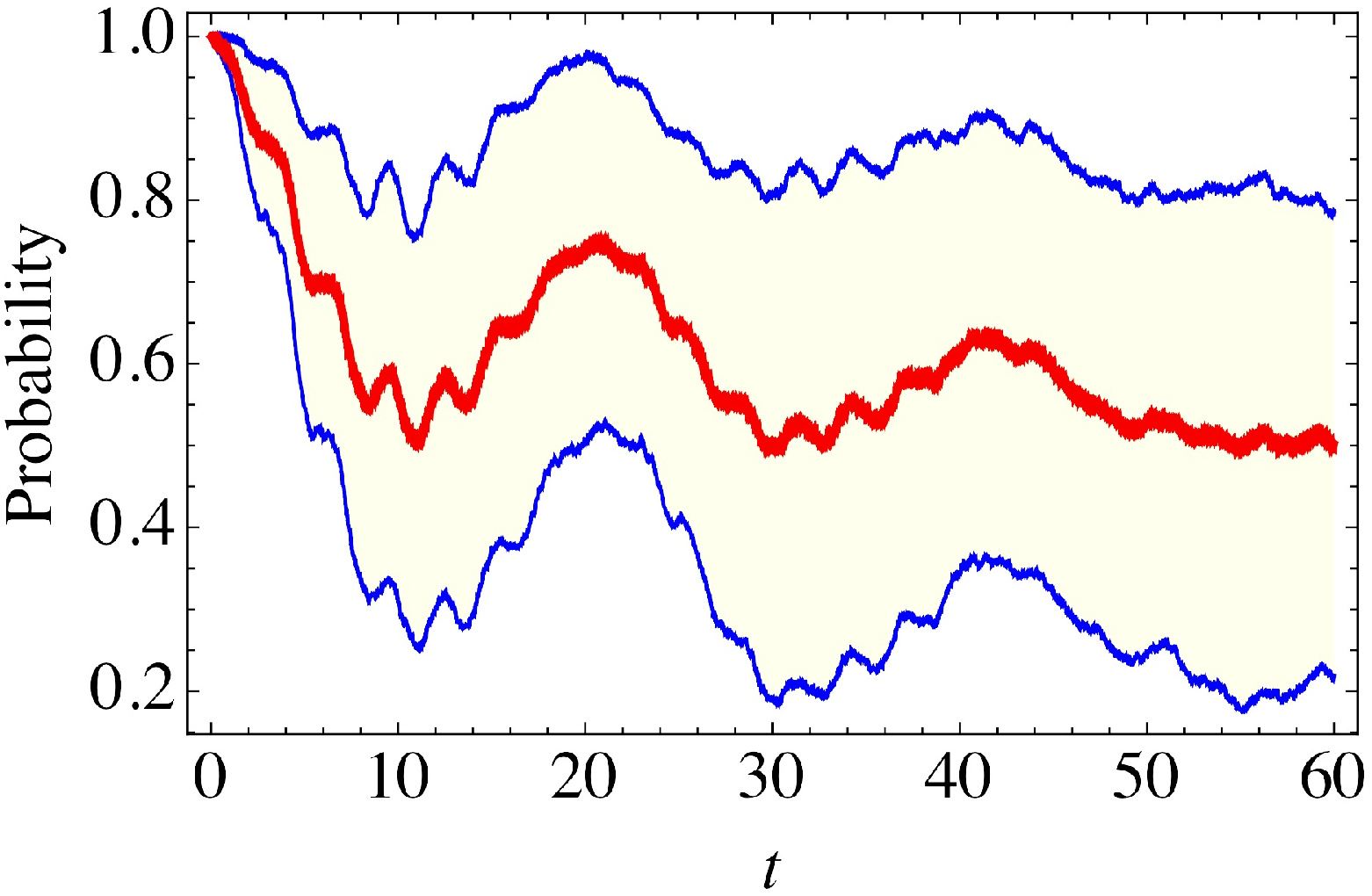}}
\centering\subfigure[]{\includegraphics[width=0.3\textwidth]
{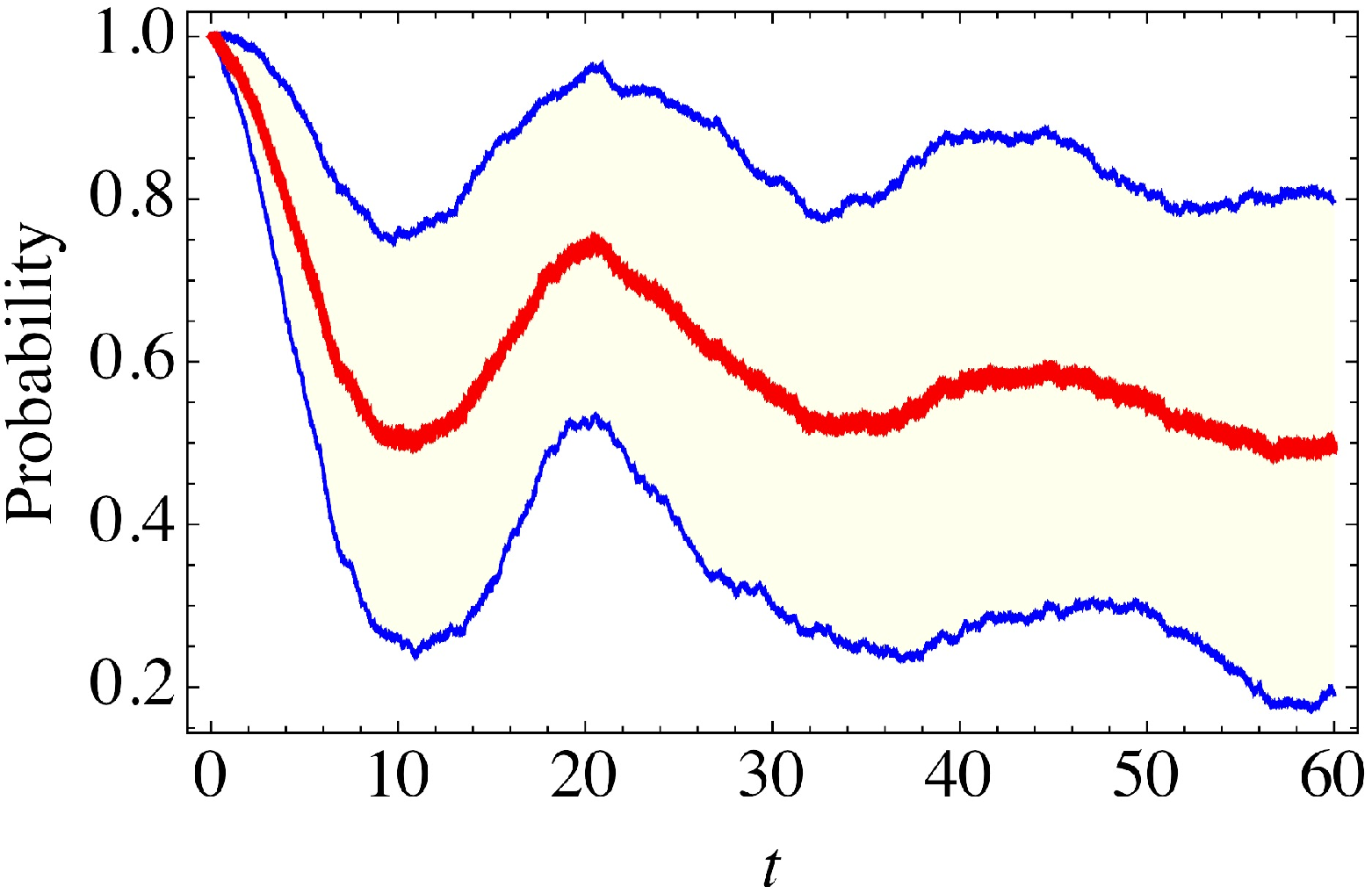}}
\centering\subfigure[]{\includegraphics[width=0.3\textwidth]
{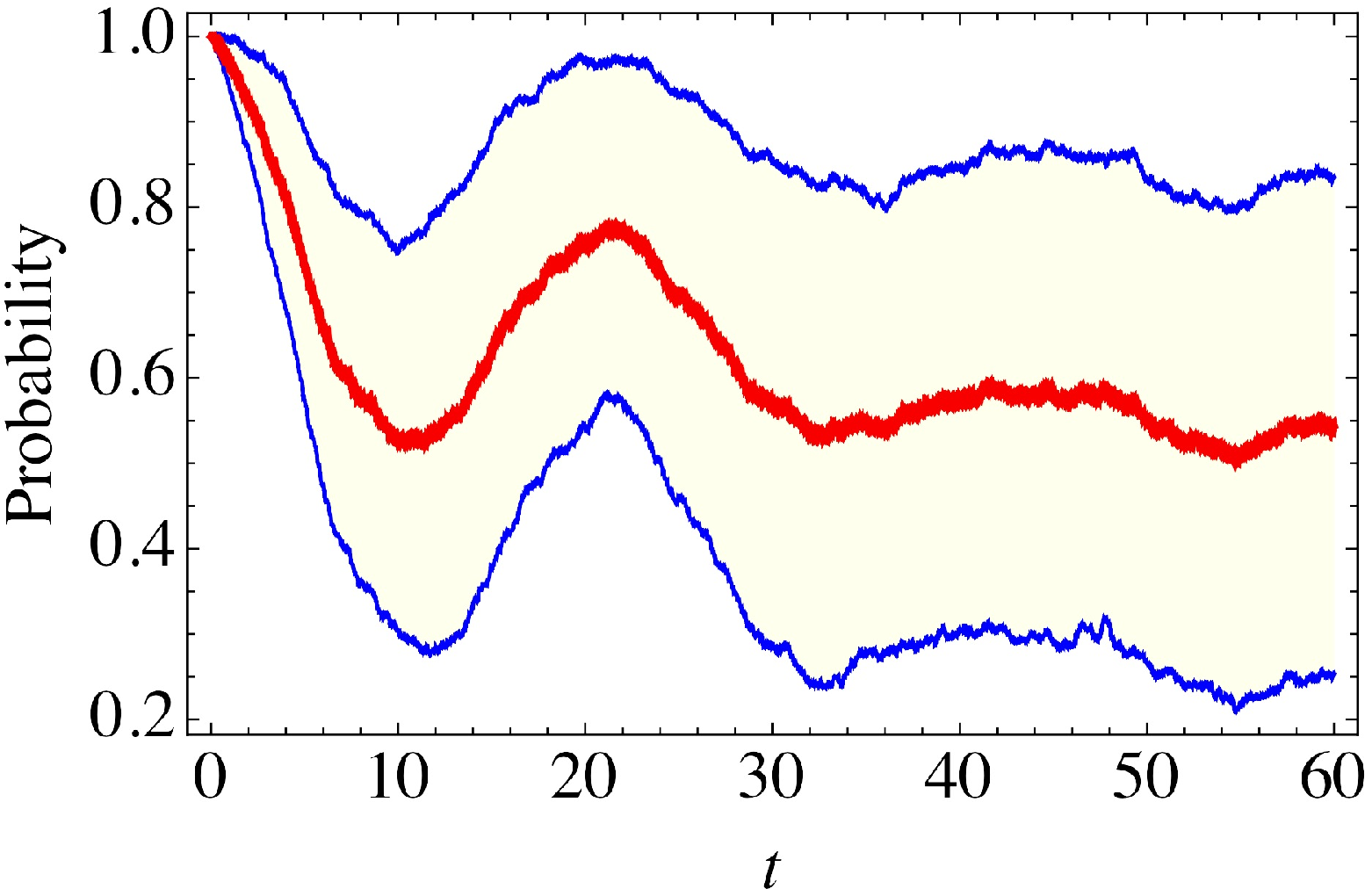}} 
\caption{(Color online) Decoherence and dephasing of off-resonance
transitions.  (a) Average probability $\overline{P_b(t)} =
\overline{\psi_b^*(t) \psi_b(t)}$ versus time for $\delta=1.2$,
$\omega=1.0$, $\Omega=0.2$ and volatilities $w_{0,1} = w_{0,2} =
w_{0,3} = 0.1$.  (b) Same as (a) (i.e., $\Delta=0.2$), except
calculated using the RWA. (c) Same as (b), except that the
non-commutativity of the RWA transformation and the transverse
stochastic Hamiltonian terms is not properly accounted for.}
\label{Fig_stochastic_nonRWA_d_1.2_W_0.2}
\end{figure}

The RWA (i.e., the transformation to the rotating frame) for the
Schr\"{o}dinger equation in Eq.~(\ref{Schr_Langevin_gen'}) [or
(\ref{Schr_Langevin_gen})] must be carried out with care because the
unitary transformation matrix in Eq.~(\ref{Eq:RWA_trans}) does not
commute with the $\sigma_x$ and $\sigma_y$ stochastic terms in
(\ref{Schr_Langevin_gen'}).  The transformation of the Gaussian white 
noise Hamiltonian in (\ref{Schr_Langevin_gen'}) gives
\begin{equation}  \label{Eq:trans_stoch_H}
    \breve{\cal H}_{\mathrm{st}}(t) = {\cal U}^\dag (t) \, [w_0 
    {\boldsymbol \xi} \cdot {\boldsymbol \sigma}] \, {\cal U}(t) = w_0
    \begin{pmatrix} \xi_z & i \, e^{-i\omega t} (\xi_x - i \xi_y) \\
    -i \, e^{i\omega t} (\xi_x + i \xi_y) & -\xi_z \end{pmatrix} ,
\end{equation}
where ${\cal U}(t)$ is given in Eq.~(\ref{Eq:RWA_trans}), and the 
SLSDE RWA for Gaussian white noise becomes
\begin{equation}  \label{Schr_Langevin_gen''}
    {\dot \psi} = -i \, \left({\cal H}_{\mathrm{RWA}} - \frac{3
    w_0^2}{2} {\bf 1} + \breve{\cal H}_{\mathrm{st}}(t) \right) \psi .
\end{equation}
Figure~\ref{Fig_stochastic_nonRWA_d_1_W_0.2}(b) shows the average
probability $\overline{P_b(t)}$, and the average plus/minus the
standard deviation of the probability, calculated using
Eq.~(\ref{Schr_Langevin_gen''}).
Figure~\ref{Fig_stochastic_nonRWA_d_1_W_0.2}(c) shows the results of
using $i {\dot \psi} = \left({\cal H}_{\mathrm{RWA}} - \frac{3
w_0^2}{2} {\bf 1} + {\cal H}_{\mathrm{st}}(t) \right) \psi$, which
neglects the fact that the RWA transformation and the transverse
stochastic Hamiltonian do not commute.  There is little difference
between the results in Figs.~\ref{Fig_stochastic_nonRWA_d_1_W_0.2}(b)
and (c), which is not surprising, given that the time dependence of
the harmonic function $e^{i\omega t}$, i.e., $\omega^{-1}$, is slow
compared to the correlation time $\tau_{\mathrm{corr}}$ of white noise, 
which is effectively zero (i.e., infinitesimal).  A significant difference 
will occur only if $\omega \tau_{\mathrm{corr}}
\gtrapprox 1$.  Only for noise with a correlation time
$\tau_{\mathrm{corr}}$ comparable to $\omega^{-1}$ or larger are large
differences are expected.  In Sec.~\ref{Sec:OU} we discuss the case of
an Ornstein--Uhlenbeck process with mean reversion rates comparable to
the frequency $\omega$; for such cases, we expect a substantial
difference between the results of taking the non-commutation into
account or not.
Figures~\ref{Fig_stochastic_nonRWA_d_1.2_W_0.2}(a)-(c) show the
off-resonance case, $\delta=1.2$, $\omega=1.0$, $\Omega=0.2$.  Again,
here there is very little difference between the results in
Figs.~\ref{Fig_stochastic_nonRWA_d_1.2_W_0.2}(b) and (c), for the same
reasons just discussed.

\section{Master (Liouville--von Neumann) Equation Results}
\label{Sec:Master_Eqs}

\begin{figure}
\centering\subfigure[]{\includegraphics[width=0.3\textwidth]
{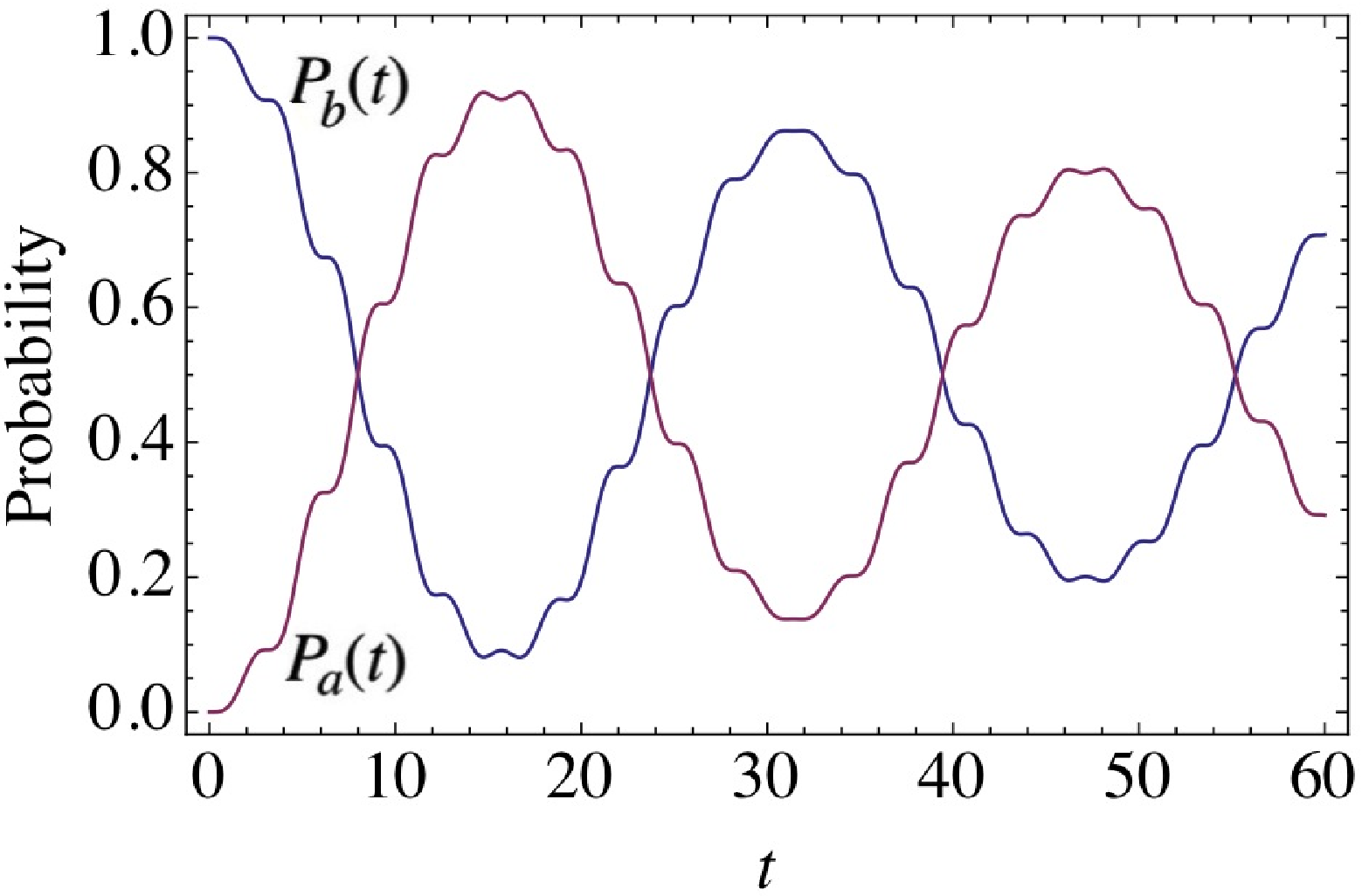}}
\centering\subfigure[]{\includegraphics[width=0.3\textwidth]
{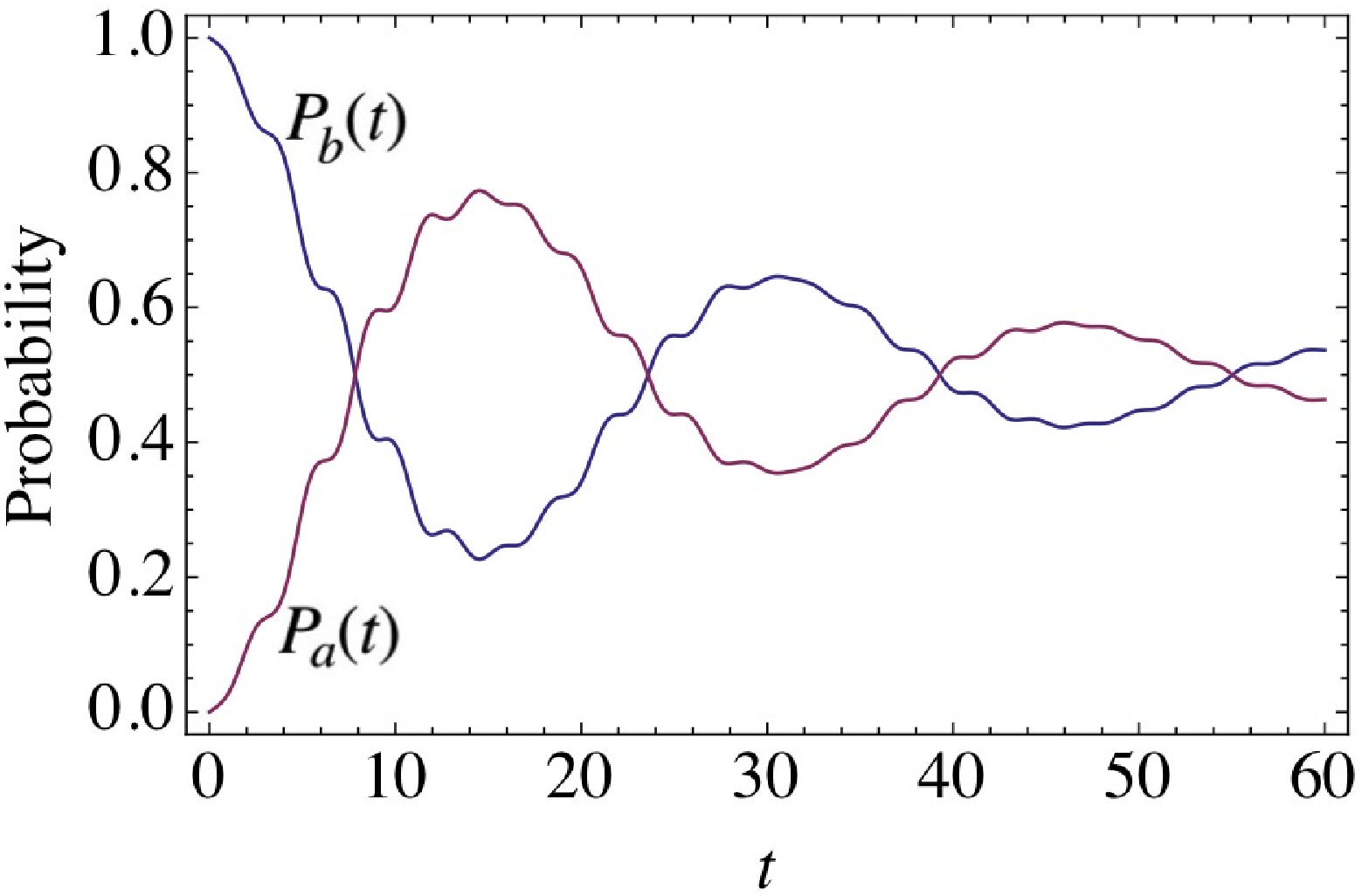}}
\centering\subfigure[]{\includegraphics[width=0.3\textwidth]
{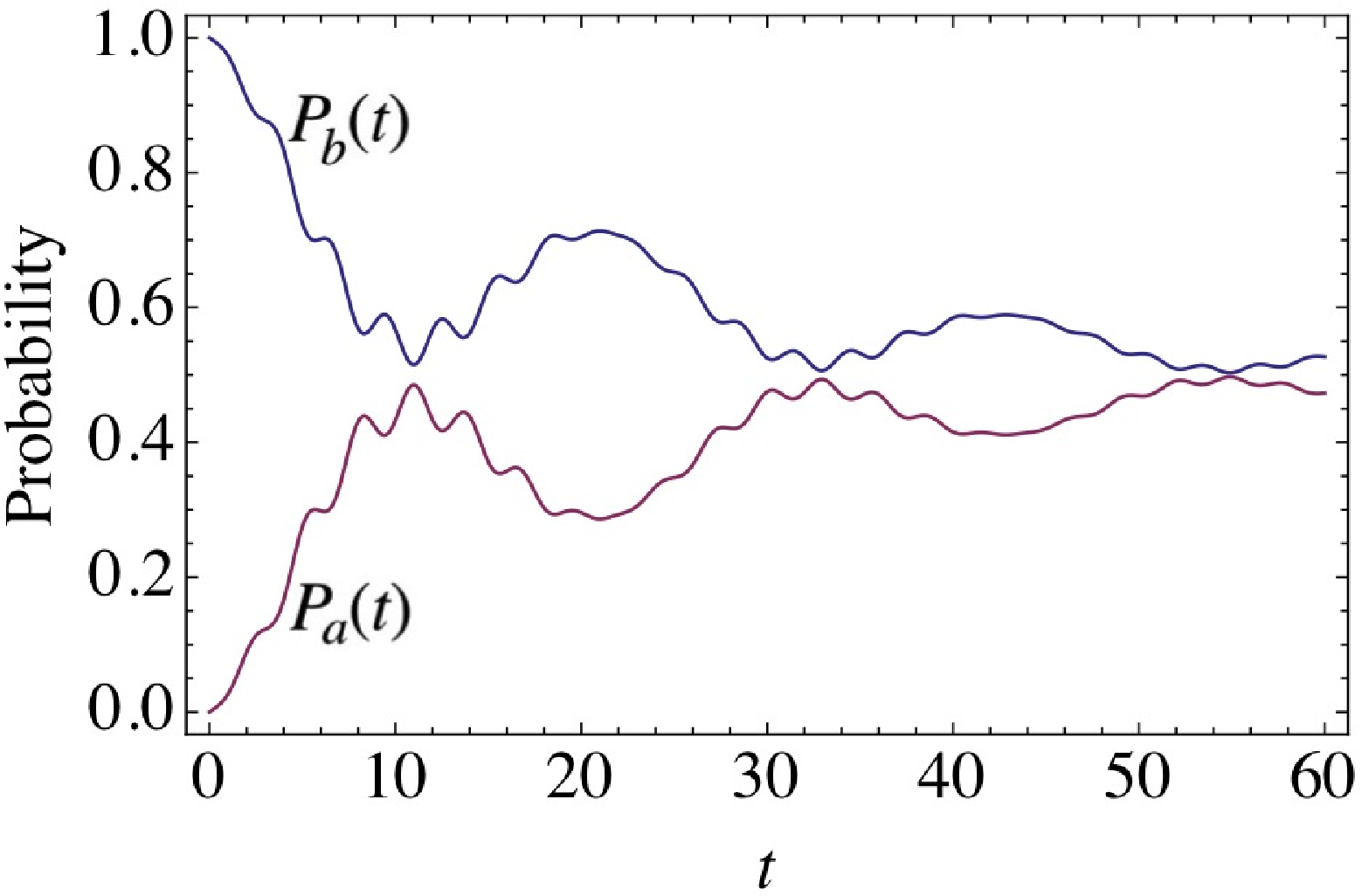}} 
\caption{(Color online) Master (Liouville-von Neumann density matrix)
equation calculations of dephasing and decoherence.  (a) Dephasing of
the probability $\rho_{bb}(t)$ versus time for the on-resonance case,
$\delta=1.0$, $\omega=1.0$, $\Omega=0.2$, and white noise in the $z$
component of the magnetic field with volatility $w_{0} = 0.1$.  (b)
Decoherence and dephasing of the probability $\rho_{bb}(t)$ versus
time for $\delta=1.0$, $\omega=1.0$, $\Omega=0.2$, and isotropic white
noise with volatilities $w_{0,1} = w_{0,2} = w_{0,3} = 0.1$.  (c) Same
as for (b), except for the off-resonance case with $\delta=1.2$,
$\omega=1.0$.}
\label{Fig_DM_nonRWA}
\end{figure}

As already mentioned, white noise gives average results that are 
identical to those obtained with a Markovian Liouville-von
Neumann density matrix equation having Lindblad terms.  
For the isotropic white noise case in Eq.~(\ref{Schr_Langevin_gen'})
the corresponding density matrix equation is
\begin{equation}  \label{Eq:DM_gen'}
    {\dot \rho} = -i [{\cal H}(t), \rho(t)] + w_0^2 \left(3 \rho(t) -
    \sum_i \sigma_i \, \rho(t) \, \sigma_i \right) .
\end{equation}
Figure~\ref{Fig_DM_nonRWA} shows the results of such
density matrix calculations.  Figure~\ref{Fig_DM_nonRWA}(a) is for
the on-resonance dephasing case with Lindblad operator $\sigma_z$
[this case of white noise only in the $z$ component of the magnetic 
field, see Eq.~(\ref{Eq:H_stochastic_z}), gives the master equation,
${\dot \rho} = -i [{\cal H}(t), \rho(t)] + w_0^2 (\rho(t) - 
\sigma_z \, \rho(t) \, \sigma_z)$],
Fig.~\ref{Fig_DM_nonRWA}(b) is for the on-resonance isotropic white
noise case, and Fig.~\ref{Fig_DM_nonRWA}(c) is for the off-resonance
case with $\Delta=1.2$, $\omega=1.0$.  In particular, $\rho_{bb}(t)$
using the density matrix (master equation) treatment is identical, to
within numerical accuracy, to the average probabilities
$\overline{P_b(t)}$ computed with the stochastic differential equation
approach.  However, the Liouville-von Neumann density matrix approach 
cannot easily determine the distribution ${\cal D}[P_b(t)]$ of the 
probability $P_b(t)$ (to do so would require calculating $P_b^m(t)$ 
for all poowers $m$), which can be directly obtained using the 
stochastic equation approach.

\begin{figure}
\centering\subfigure[]{\includegraphics[width=0.4\textwidth]
{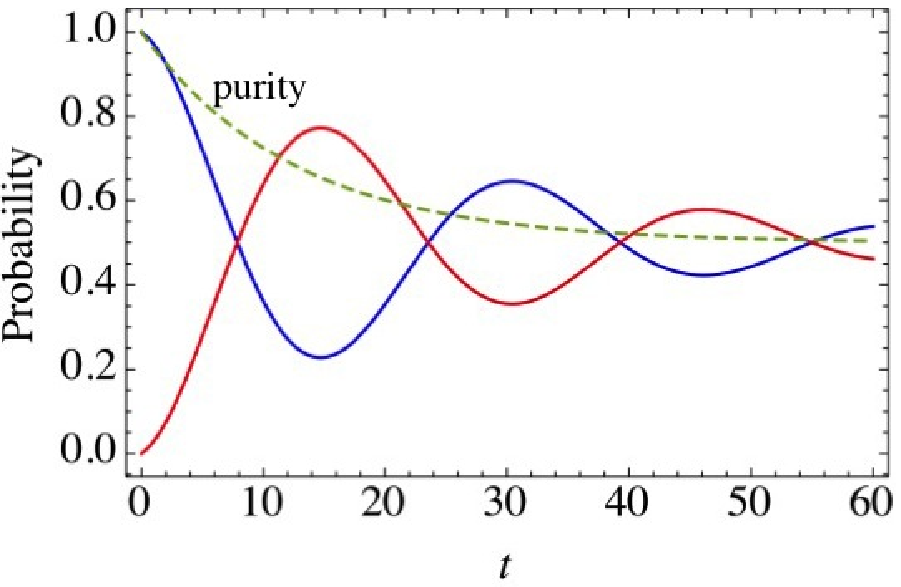}}
\centering\subfigure[]{\includegraphics[width=0.4\textwidth]
{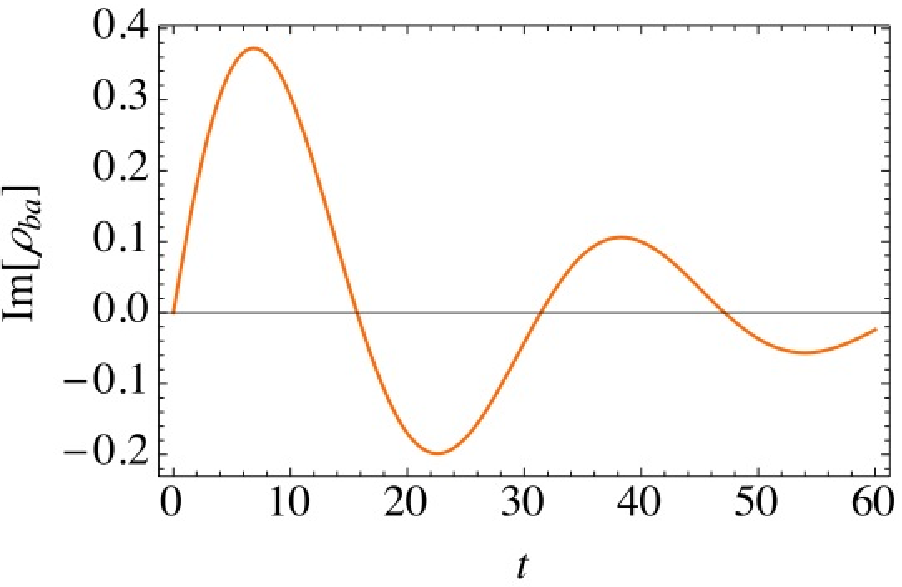}}
\caption{(Color online) Analytic results obtained using
Eq.~(\ref{eq:rho_an}) for the density matrix for isotropic white noise
in the on-resonance case with $\delta=1.0$, $\omega=1.0$,
$\Omega=0.2$, and $w_0 \equiv w_{0,1} = w_{0,2} = w_{0,3} = 0.1$.  (a)
Density matrix elements $\rho_{bb}(t)$ and $\rho_{aa}(t)$, and the
purity ${\mathrm{Tr}}[\rho^2(t)]$ (dashed green curve) versus time.
Compare with the non-RWA results in Fig.~\ref{Fig_DM_nonRWA}(b).  (b)
Off-diagonal density matrix $\rho_{ba}(t)$ versus time.  The real part
of $\rho_{ba}(t)$ vanishes for the on-resonance case.}
\label{Fig_DM_RWA}
\end{figure}

Now, consider the RWA.  The analytic solution to Eq.~(\ref{Eq:DM_gen'})
(isotropic Gaussian white noise), using the RWA Hamiltonian
${\cal H}_{\mathrm{RWA}}(t)$ instead of ${\cal H}(t)$, is given by
\begin{eqnarray}
&& \rho_{bb}(t) = \frac{1}{2}\left\{1 + \frac{e^{-4 w_0 t}[\Delta^2 + 
\Omega^2 \cos(\Omega_g t)]}{4 \Omega_g^2}\right\} , \nonumber \\
&& \rho_{ba}(t) = \frac{e^{-4 w_0 t} \Omega \left[\Delta - 
\Delta \cos(\Omega_g t)  + i \Omega_g \sin(\Omega_g t) \right]}
{4 \Omega_g^2} , \nonumber \\
&& \rho_{ab}(t) = \rho_{ba}^*(t)  , \nonumber \\
&& \rho_{aa}(t) = \frac{1}{2}\left\{1 - \frac{e^{-4 w_0 t}[\Delta^2 + 
\Omega^2 \cos(\Omega_g t)]}{4 \Omega_g^2}\right\} .
\label{eq:rho_an}
\end{eqnarray}
where $\Omega_g \equiv \sqrt{\Omega^2 + \Delta^2}$.  
Figure~\ref{Fig_DM_RWA}(a) plots the
probabilities $\rho_{bb}(t)$ and $\rho_{aa}(t)$ and the purity
${\mathrm{Tr}}[\rho^2(t)]$ versus time for the on-resonance case.
As $t \to \infty$, the purity goes to 1/2 and the density matrix 
decays to the democratic state $\frac{1}{2} \binom{1 \, 0} {0 \, 1}$.
Figure~\ref{Fig_DM_RWA}(b) plots the coherence $\rho_{ba}(t)$ versus 
time; it has only an imaginary component and it decays to zero as 
$t \to \infty$.  The decay rate of the 
populations and the coherence is $4 w_0$, as is evident from the 
expressions in Eq.~(\ref{eq:rho_an})].  Properly accounting for 
the non-commutativity of the RWA transformation and the
stochastic Hamiltonian, i.e., using the stochastic Hamiltonian in
Eq.~(\ref{Eq:trans_stoch_H}), does not significantly affect the
numerical results for white Gaussian noise.  The full RWA
probabilities, including non-commutativity effects, are
indistinguishable by eye from the results shown in
Fig.~\ref{Fig_DM_RWA}.  The full RWA coherence 
${\mathrm{Im}}[\rho_{ba}(t)]$ is
also indistinguishable by eye, and the ${\mathrm{Re}}[\rho_{ba}(t)]$
is more than an order of magnitude smaller than the imaginary part.

\section{Ornstein--Uhlenbeck Process}  \label{Sec:OU}

Many kinds of stochastic processes have been studied.  In order to see
significant effects due to the non-commutativity of the RWA transformation
and stochastic part of the total Hamiltonian $\tilde H(t) = {\cal
H}(t) + {\cal H}_{\mathrm{st}}(t)$, we need a stochastic process with
a correlation time $\tau_{\mathrm{corr}}$ comparable to or greater
than the timescale of the driven two level system $\omega^{-1}$.  A
well-known finite-correlation-time stochastic process is the
Ornstein--Uhlenbeck process, which is an example of Gaussian colored noise,
which is a generalization of Brownian motion \cite{OU_30}.  The mean 
and the autocorrelation function of an Ornstein--Uhlenbeck process are
\begin{equation}  \label{OU_noise}
    \overline{{\cal O}_i(t)} = {\cal O}_{0,i} \, e^{-\vartheta_i t} + \mu_i (1 -
    e^{-\vartheta_i t}) , \quad \overline{{\cal O}_i(t) \, {\cal O}_j(t')} =
    \delta_{ij} \frac{w_{0,i}^2}{2\vartheta_i} e^{-\vartheta_i (t+t')}
    [e^{\vartheta_i \, {\mathrm{min}}(t,t')} - 1] .
\end{equation}
Here $\vartheta_i$ is the mean reversion rate of the
Ornstein--Uhlenbeck process ${\cal O}_i(t)$, which is the inverse of the
noise correlation time, $\tau_{\mathrm{corr},i} = \vartheta_i^{-1}$,
$w_{0,i}$ is its volatility, and $\mu_i$ is the mean of the process,
which we take to vanish, $\mu_i = 0$; we also take ${\cal O}_{0,i} = 0$.
The stochastic differential equations that we solve are,
\begin{subequations}  \label{SL_stoch_OU}
\begin{equation}  \label{SL1_stoch_OU}
    d \psi(t) = \left[ -i \, {\cal H} \psi + \sum_i {\cal O}_i(t) \sigma_i
    \psi(t) \right] \, dt ,
\end{equation}
\begin{equation}  \label{SL2_stoch_OU}
    d {\cal O}_i(t) = \vartheta_i \left[\mu_i - {\cal O}_i(t) \right]\, dt +
    w_{0,i} \, dw_i(t) .
\end{equation}
\end{subequations}
For determining the effects of the non-commutativity of
the RWA transformation and stochastic terms in the total Hamiltonian,
it is sufficient to use only the term $i = x$ in the sum in
Eq.~(\ref{SL1_stoch_OU}).  Doing so simplifies the convergence of the
calculation relative to using isotropic Ornstein--Uhlenbeck noise.

\begin{figure}
\centering\subfigure[]{\includegraphics[width=0.3\textwidth]
{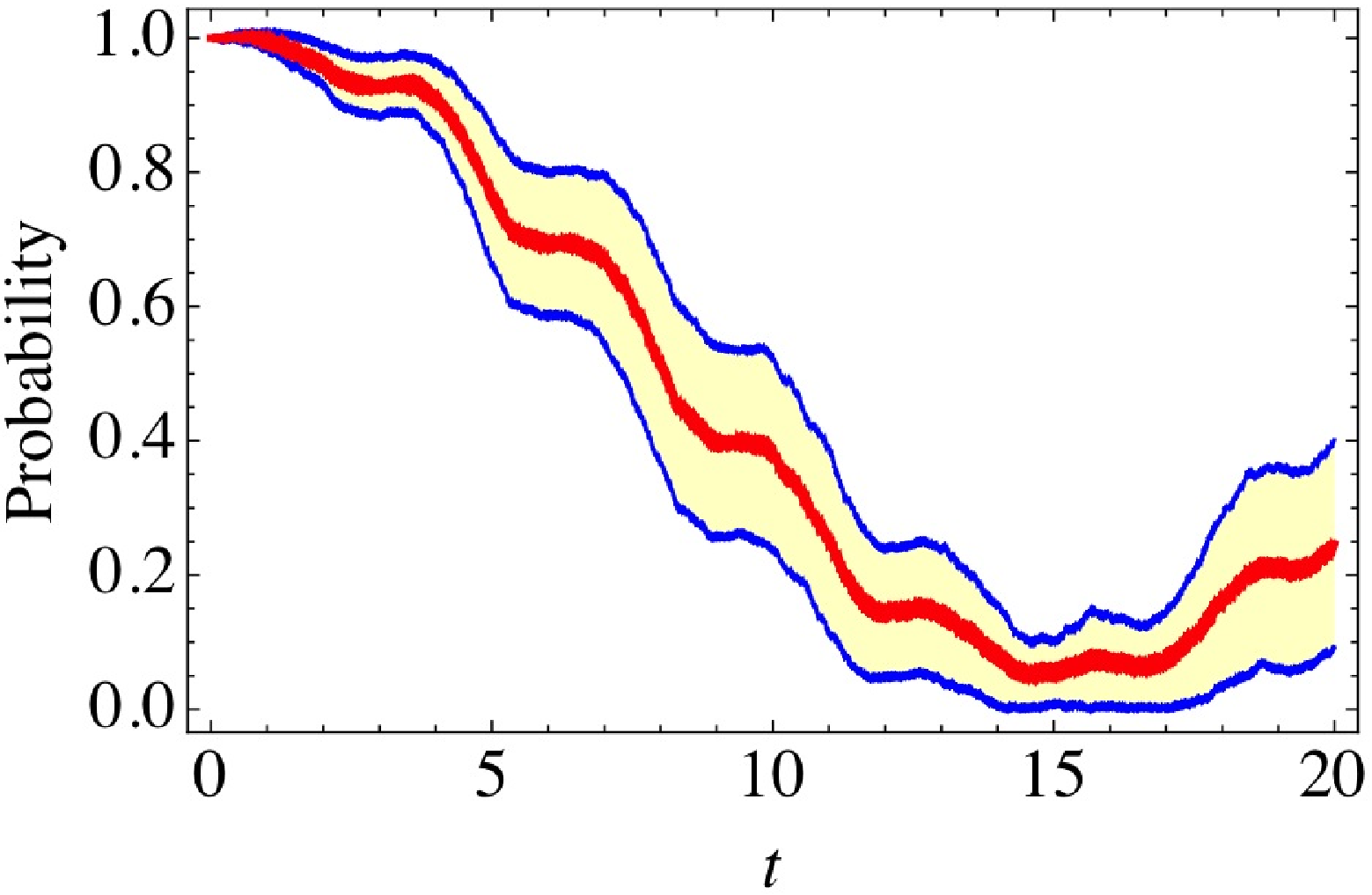}} 
\centering\subfigure[]{\includegraphics[width=0.3\textwidth]
{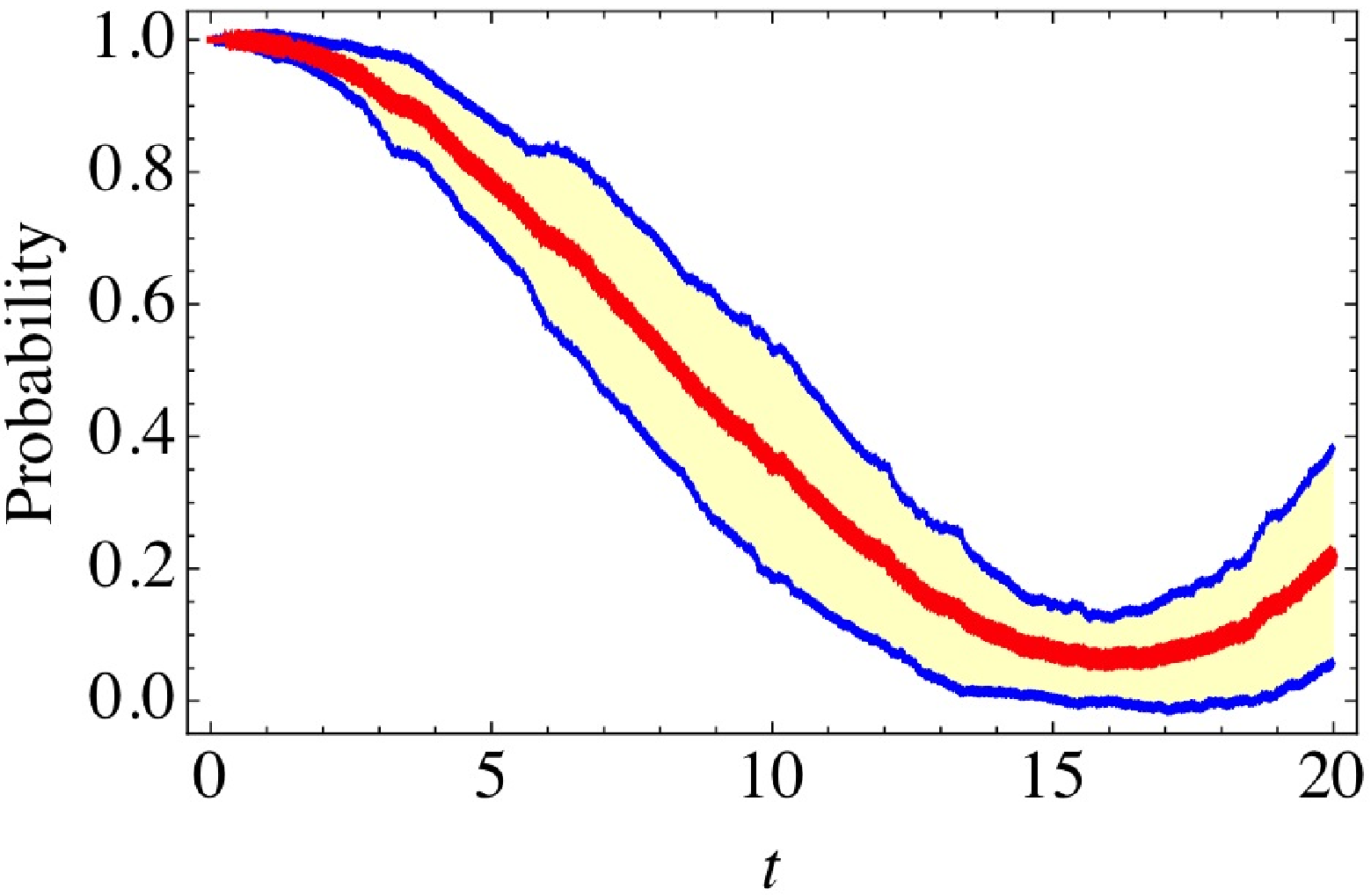}}
\centering\subfigure[]{\includegraphics[width=0.3\textwidth]
{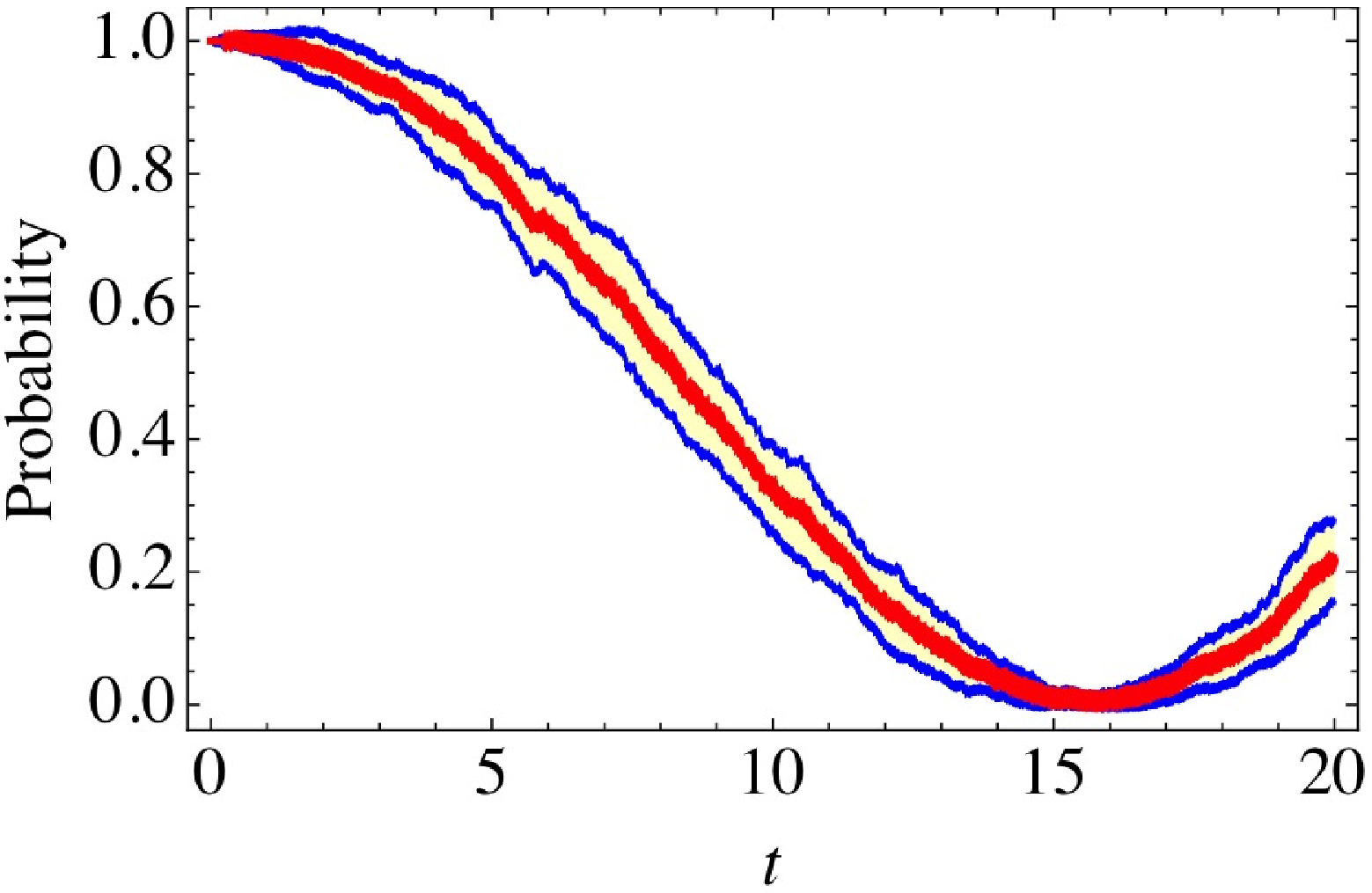}}
\caption{(Color online) Ornstein-Olhenbeck noise and
on-resonance radiation, highlighting the effects of the 
non-commutativity of the RWA transformation and the
transverse stochastic Hamiltonian.  (a) Probability
$\overline{P_b(t)}$ versus time for the on-resonance case,
$\delta=1.0$, $\omega=1.0$ (so $\Delta = 0$), $\Omega=0.2$, and
Ornstein-Olhenbeck noise in the $x$ component of the magnetic field,
with volatility $w_{0,x} = 0.1$ and mean reversion rate $\vartheta_x =
1$ (so $\omega \tau_{\mathrm{corr}} = 1$).  (b) Same as (a) except
calculated using the RWA. (c) Same as (b), except that the
non-commutativity of the RWA transformation and the transverse
stochastic Hamiltonian term is not properly accounted for.}
\label{Fig_OU_nonRWA_sx}
\end{figure}

\begin{figure}
\centering\subfigure[]{\includegraphics[width=0.3\textwidth]
{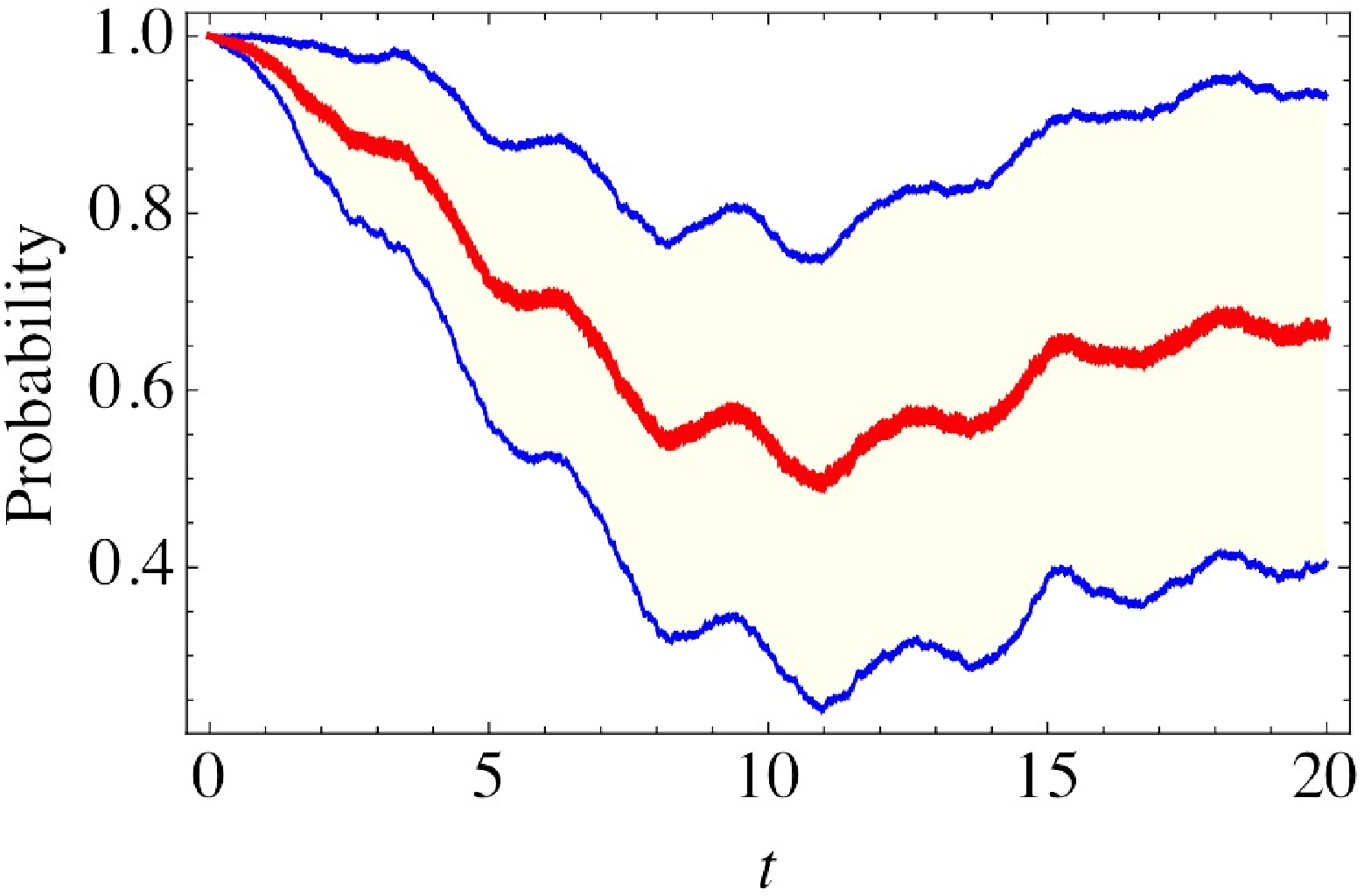}} 
\centering\subfigure[]{\includegraphics[width=0.3\textwidth]
{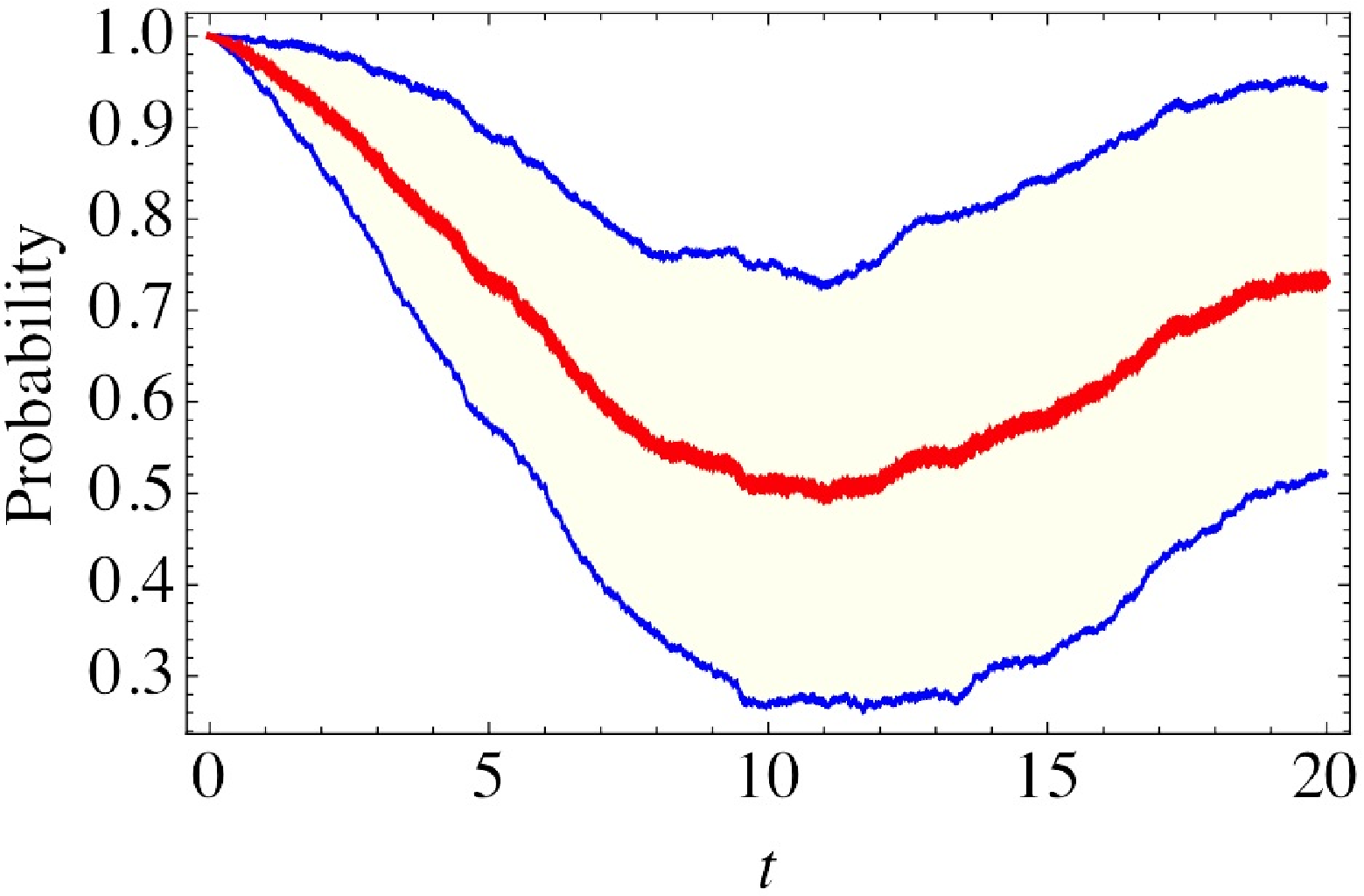}}
\centering\subfigure[]{\includegraphics[width=0.3\textwidth]
{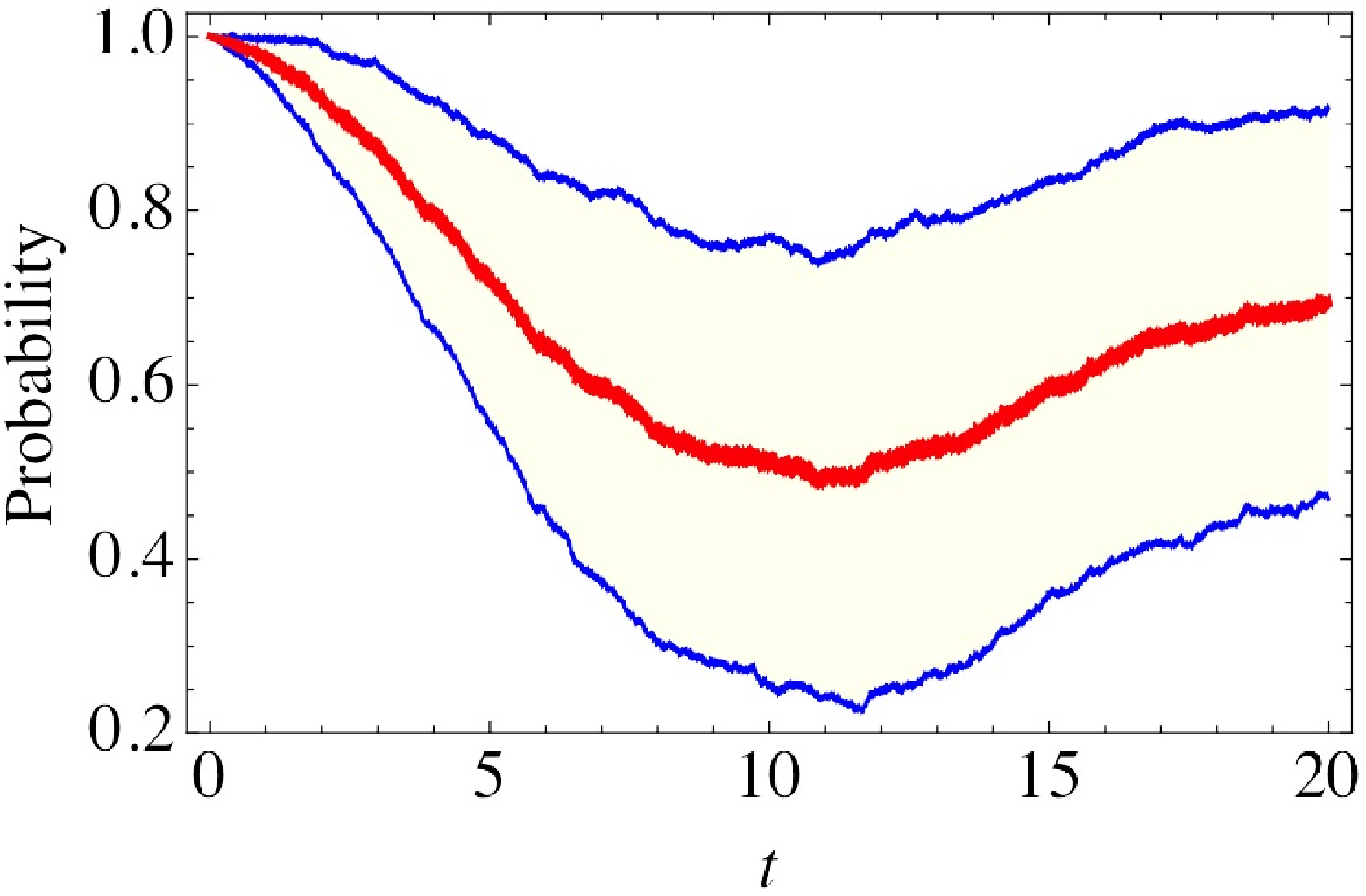}}
\caption{(Color online) Isotropic Ornstein-Olhenbeck noise and
off-resonance radiation, $\Delta = 0.2$.  (a) Probability
$\overline{P_b(t)}$ versus time for the off-resonance case,
$\delta=1.2$, $\omega=1.0$ (so $\Delta = 0.02$), $\Omega=0.2$, and
isotropic Ornstein-Olhenbeck noise in the magnetic field, with 
volatilities $w_{0,x} = w_{0,y} = w_{0,z} = 0.1$ and mean reversion 
rate $\vartheta_x = 1$ (so $\omega \tau_{\mathrm{corr}} = 1$).  
(b) Same as (a) except calculated using the RWA. (c) Same as (b), 
except that the non-commutativity of the RWA transformation and 
the transverse stochastic Hamiltonian term is not properly accounted 
for.}
\label{Fig_OU_nonRWA_sxyz}
\end{figure}

Figure~\ref{Fig_OU_nonRWA_sx} shows the calculated average probability
$\overline{P_b(t)}$ versus time and the average plus and minus the 
standard deviation of the probability calculated using 
Eq.~(\ref{SL_stoch_OU}) for Ornstein--Uhlenbeck noise in the $x$ 
component of the magnetic field for the on-resonance case, $\delta=1.0$, 
$\omega=1.0$, $\Omega=0.2$.  Figure~\ref{Fig_OU_nonRWA_sx}(b) is calculated 
using the RWA, and for comparison purposes only, (c) shows the results
using a RWA but ignoring the non-commutativity of 
the RWA transformation and the transverse stochastic Hamiltonian term, 
i.e., ignoring the factors $e^{\pm i \omega t}$ in the off-diagonal
elements of the Hamiltonian 
\begin{equation} \label{Eq:H_RWA}
    \tilde {\cal H}_{\mathrm{RWA}}(t) = {\cal H}_{\mathrm{RWA}} + 
    {\cal H}_{\mathrm{st,RWA}}(t) = \begin{pmatrix} -\Delta & \Omega \\
    \Omega & 0 \end{pmatrix} + \begin{pmatrix} b_z(t) & 
    i \, e^{-i \omega t} [b_x(t) - i b_y(t)] \\
    -i \, e^{i \omega t} [b_x(t) + i b_y(t)] & - b_z(t) \end{pmatrix} .
\end{equation}
The oscillating factors $e^{\pm i \omega t}$ in the off-diagonal terms
can be ignored if $\omega \tau_{\mathrm{corr}} \ll 1$, but not
otherwise.  In Fig.~\ref{Fig_OU_nonRWA_sx} we used $\omega
\tau_{\mathrm{corr}} = 1$, so we expect that the non-commutativity
cannot be ignored, and we took noise only in the $x$-component of
the magnetic field.  The calculations in Fig.~\ref{Fig_OU_nonRWA_sx}
were hard to converge with respect to the time-step used, hence we
only continued them out to a final time of $t = 20$.  Note that the
standard deviation in Fig.~\ref{Fig_OU_nonRWA_sx}(c) is significantly
reduced relative to (a) and (b), and the width becomes very close to
zero at $t = 15.5$ where the average probability goes to zero, unlike
the results in (a) and (b).  Clearly, the results of ignoring the
non-commutativity of the RWA transformation and the transverse
stochastic Hamiltonian are very different from the RWA taking the
non-commutativity into account.  The minimum of the probability in (c)
is shifted to somewhat smaller time and is {\em much} closer to
zero probability than in (b); moreover the standard deviation in (c)
is much smaller than in (b).  We also expect a difference between
taking and not taking the non-commutativity into account in a master
equation approach.  The master equation for OU noise could in 
principle be determined using cumulant generating functional methods 
and requires calculation of time-ordered exponential functions 
\cite{Kubo_cumulants} but this is a difficult task.
Figure~\ref{Fig_OU_nonRWA_sxyz} shows $\overline{P_b(t)}$ for
isotropic Ornstein--Uhlenbeck noise for the off-resonance case,
$\Delta = 0.2$.  Here, the differences between (b) and (c) are
small.  No difference between (b) and (c) results due to the 
$z$-component of the noise field, whose noise Hamiltonian commutes 
with the RWA transformation; moreover, there is some compensation 
which takes place between the $x$ and $y$ components.

\section{Summary and conclusions}  \label{Sec:Summary}

Much of our experience with quantum dynamics comes from
applying it to two-level systems driven by an electromagnetic field.
But such systems are never truly isolated, and their interaction with
their environment affect their mysterious quantum properties, i.e.,
their quantum coherence.  Such interaction is at the heart of the
fundamental limitations of quantum metrology and quantum information
processing.  Using the Schr\"{o}dinger-Langevin stochastic differential 
equation (SLSDE) formalism, we studied the dynamics of a two-level
quantum system driven by single frequency electromagnetic field, with
and without making the rotating wave approximation (RWA).  If the 
transformation to the rotating frame does not commute with the stochastic 
Hamiltonian, i.e., if the stochastic field has nonvanishing components in 
the transverse direction, the RWA modifies the stochastic terms 
in the Hamiltonian.  The decay terms in a master equation (i.e., the 
Liouville--von Neumann density matrix equation) will also be affected.  
We found that for Gaussian white noise, the master equation for the density 
matrix is easily derived from the SLSE, with and without the RWA. For the 
RWA, both the SSLE and the derived master equation have Lindblad terms 
with coefficients that are time-dependent (i.e., the master 
equation is time-local \cite{Fleming_12b}) when the 
non-commutation of the RWA transformation and the noise Hamiltonian is 
properly accounted for.  But since $\omega \tau_{\mathrm{corr}}$ 
effectively vanishes for white Gaussian  noise, it is not necessary to 
take the non-commutation into account, independent of the strength of the 
noise ($w_0$), and we obtain an analytic expression for the density matrix 
of the system, Eq.~(\ref{eq:rho_an}), which fully describes the dynamics 
of the two-level system in the presence of the noise. On the other hand, 
for the non-Markovian Ornstein--Uhlenbeck noise case, the RWA dynamics 
must be calculated taking the non-commutation of the RWA transformation 
and the noise Hamiltonian into account when $\omega \tau_{\mathrm{corr}} 
\gtrapprox 1$.

Decoherence and dephasing of two-level systems can be probed by
measuring the population decay ($T_1$) and the transverse relaxation 
time ($T_2$) in magnetic resonance studies \cite{Jarmola_12, Kehayias_14}.  
One well-studied physical system in which such studies have been
carried out is the negatively-charged nitrogen-vacancy (NV) color center 
in diamond.  An NV center consists of a substitutional nitrogen atom 
adjacent to a missing carbon atom within the diamond crystal lattice 
\cite{Doherty_13}.  The negatively charged NV center has a
discrete electronic energy level structure and a ground electronic
state of symmetry ${}^3 A_2$, where this state designation
refers to an irreducible representation of the $C_{3v}$ group.  The
three electronic magnetic sub-levels of the triplet ground state are
$| S, M \rangle$, where $S = 1$ and $M = 0, \pm 1$, with the $z$ axis
(quantization axis) taken along the NV axis.  The three $S = 1$ ground
state levels are split by a spin-spin (crystal field) interaction that
raises the energy of the $| 1, \pm 1 \rangle$ states with respect to
the $| 1, 0 \rangle$ state by ${\cal D} = 2.87$ GHz.  The NV system
can behave like a two-level system if one of the three states is not
allowed to be populated.  The main sources of decoherence are from the
paramagnetic impurity spin bath, which dominates at high nitrogen
concentration, and interactions with the spin 1/2 ${}^{13}$C nuclei
\cite{Balas_09, Taylor_08}.  Population decay, $T_1$, is dominated by
Raman-type interactions with lattice phonons at high temperature (room
temperature and above), Orbach-type interaction with local phonons at
lower temperatures \cite{Jarmola_12, Takahashi_08, Redman_91}, and at
temperatures below about 100 K, density-dependent cross-relaxation effects 
between NV centers and between NVs and other impurities.  At these low 
temperatures, the resulting $T_1$ can be dramatically tuned using an external 
magnetic field \cite{Jarmola_12}.  For dilute samples, the contribution of
NV--NV dipolar interactions to the magnetic resonance broadening can
be approximated by assuming that each NV center couples to 
neighboring NV centers, to substitutional nitrogen (P1) centers, which 
have a spin of 1/2, and with ${}^{13}$C nuclei, which have a nuclear 
spin of 1/2 and a natural abundance of about 1\%.  Dipolar
coupling with other NV centers leads to a spin-relaxation contribution
on the order of $\gamma_{\mathrm{NV}} \approx (g_s \mu_B)^2
n_{\mathrm{NV}}$, where $n_{\mathrm{NV}}$ is the NV concentration
\cite{Taylor_08, Acosta_09}.  For [NV] = 15 ppm, this corresponds to
$\gamma_{\mathrm{NV}} \approx 10^6$ s${}^{-1} \approx \gamma_{C}$,
where $\gamma_{C}$ is the spin relaxation rate to to the ${}^{13}$C
nuclei.  Since the dynamics of the ${}^{13}$C nuclear spin is slow,
it can be modeled, to good approximation, as quasi-static Guassian noise.
Since the spin dynamics of the NV centers and PI centers in diamond
are fast, the contribution of NV--NV and NV--P1 interactions can 
be modeled, to good approximation, as Gaussian white noise.  As demonstrated 
in Ref.~\cite{Kehayias_14}, CW hole-burning and lock-in detection can 
be used to eliminate the linewidth contribution 
from slowly fluctuating ${}^{13}$C nuclei while rapidly fluctuating 
magnetic fields from nearby substitutional nitrogen (P1) centers and NV 
centers continue to contribute to a reduced linewidth.  Hence, by 
adjusting the external magnetic field strength and the concentrations of 
NV centers, P1 centers and ${}^{13}$C, the volatilities $w_{0}$, the 
stochastic magnetic field correlation times $\tau_{\mathrm{corr}}$ and 
the detuning from resonance $\Delta$ can be modified.  If only two of
the three triplet ground state levels $| S, M \rangle$ are populated,
the methods developed in this manuscript can be applied directly; if all 
three levels are populated, it is straightforward to generalize the spin 
1/2 treatment here to $S = 1$.  In either case, the conclusions we 
obtained are quite general and are expected to apply to the NV diamond 
system.  One would, of course need to know the correlation times and the 
strength of the noises affecting the NV centers.  
Specifically, if the correlation time $\tau_{\mathrm{corr}}$ 
of the noise (of the bath coupled to the system) is of order of the 
frequency of the electromagnetic field $\omega$ that couples the levels, 
the non-commutativity of the RWA transformation and the noise Hamiltonian
must be taken into account, even when the criteria for validity of the RWA
for the system are satisfied.  For diamond NV centers, the resonant
frequency $\omega$ for transitions from $M = 0$ to $M = \pm 1$ is of
order GHz (with no external magnetic field, it is 2.87 GHz), so for 
$\omega \tau_{\mathrm{corr}} \approx 1$, $\tau_{\mathrm{corr}}$ must be 
of order milliseconds.  When $\omega \tau_{\mathrm{corr}} \ll 1$, the 
non-commutativity need {\em not} be taken into account.

\bigskip

{\it Acknowledgement.} This work was supported in part by grants from
the Israel Science Foundation (Grant No.  295/2011).  I am grateful to
Professors Yshai Avishai and Dmitry Budker for valuable discussions.

\bigskip

\bigskip



\begin{thebibliography}{99}
\bibitem{semiclass}
The Hamiltonian in Eq.~(\ref{Eq:H}) is often termed semiclassical
because the electromagnetic field is treated classically,
rather than using a second-quantized radiation field.
\end{thebibliography}
\end{document}